\documentclass[conference]{IEEEtran}
\ifCLASSINFOpdf
\else
\fi
\PassOptionsToPackage{table}{xcolor}
\usepackage[bookmarks=false]{hyperref}
\usepackage{graphicx}
\usepackage{float}
\usepackage{amsmath}
\usepackage{amssymb}
\usepackage{tikz}
\usepackage[binary-units=true, group-separator={,}, group-minimum-digits={3}]{siunitx}
\usetikzlibrary{shapes,arrows,positioning,calc}
\usepackage{cite}
\usepackage{arydshln}
\usepackage{xspace}
\usepackage{xurl}
\usepackage{algorithm}
\usepackage[noend]{algpseudocode}

\usepackage{wasysym}
\usepackage{tabularx}
\usepackage{xcolor}

\usepackage{booktabs}
\usepackage{tabularx}
\usepackage{multirow}
\usepackage{multicol}
\usepackage{colortbl}
\usepackage{blindtext}
\usepackage{fp}

\usepackage{pifont}
\usepackage{paralist}

\usepackage[normalem]{ulem}
\usepackage{color}
\usepackage{subcaption}

\definecolor{myblue}{HTML}{4467f5}
\usepackage[]{hyperref}
\hypersetup{
    colorlinks = true,
    linkcolor = blue,
    anchorcolor = blue,
    citecolor = blue,
    filecolor = blue,
    urlcolor = blue
}
\usepackage{minibox}
\usepackage{tcolorbox}
\newtcolorbox{mybox}[2][]{%
  colback      = gray!5!white,
  colframe     = gray!75!black,
  fonttitle    = \bfseries,
  colbacktitle =  gray!85!black,
  title        = #2,#1,
}

\definecolor{darkgreen}{RGB}{0,160,0}
\definecolor{darkred}{RGB}{178,34,34}

\colorlet{tableheadcolor}{black} 
\newcommand{\headcol}{\rowcolor{tableheadcolor}} %
\colorlet{tablerowcolor}{gray!10} 
\colorlet{tablerowcolortwo}{gray!25} 
\colorlet{tablerowcolorthree}{gray!60} 
\newcolumntype{?}{!{\vrule width 1pt}}

\newcounter{note}[section]

\colorlet{Mycolor1}{green!10!orange!90!}

\newcommand{\eg}{\textit{e.g.}\xspace}
\newcommand{\ie}{\textit{i.e.}\xspace}

\newcommand{\aka}{\textit{a.k.a.}\xspace}
\newcommand{\etal}{\textit{et al.}\xspace}

\newcommand{\major}[1]{\textcolor{black}{#1}}

\newcounter{packednmbr}

\newenvironment{packeditemize}{
\begin{list}{$\bullet$}{
\setlength{\labelwidth}{5pt}
\setlength{\itemsep}{3pt}
\setlength{\leftmargin}{\labelwidth}
\addtolength{\leftmargin}{\labelsep}
\setlength{\parindent}{0pt}
\setlength{\listparindent}{\parindent}
\setlength{\parsep}{1pt}
\setlength{\topsep}{1pt}}}{\end{list}}

\renewcommand{\paragraph}[1]{\vspace{0.02in}\noindent{\bf{#1}.}}
\newcommand{\ignore}[1]{}

\newcommand{\sysname}{\texttt{VRecKey}\xspace}

\newcommand*\myYescirc{\CIRCLE\xspace}
\newcommand*\myHalfcirc{\LEFTcircle\xspace}
\newcommand*\myNocirc{\Circle\xspace}

\begin{document}
%


\title{Non-intrusive and Unconstrained Keystroke Inference in VR Platforms via Infrared Side Channel}

\author{\IEEEauthorblockN{Tao Ni}
\IEEEauthorblockA{City University of Hong Kong\\
taoni2@cityu.edu.hk}
\and
\IEEEauthorblockN{Yuefeng Du}
\IEEEauthorblockA{City University of Hong Kong\\
yuefengdu2@cityu.edu.hk}
\and
\IEEEauthorblockN{Qingchuan Zhao}
\IEEEauthorblockA{City University of Hong Kong\\
qizhao@cityu.edu.hk}
\and
\IEEEauthorblockN{Cong Wang}
\IEEEauthorblockA{City University of Hong Kong\\
congwang@cityu.edu.hk}}

\maketitle


\begin{abstract}

Virtual Reality (VR) technologies are increasingly employed in numerous applications across various areas.
Therefore, it is essential to ensure the security of interactions between users and VR devices.
In this paper, we disclose a new side-channel leakage in the constellation tracking system of mainstream VR platforms, where the infrared (IR) signals emitted from the VR controllers for controller-headset interactions can be maliciously exploited to reconstruct unconstrained input keystrokes on the virtual keyboard non-intrusively.
We propose a novel keystroke inference attack named \sysname to demonstrate the feasibility and practicality of this novel infrared side channel.
Specifically, \sysname leverages a customized 2D IR sensor array to intercept ambient IR signals emitted from VR controllers and subsequently infers \textit{(i)} character-level key presses on the virtual keyboard and \textit{(ii)} word-level keystrokes along with their typing trajectories.
We extensively evaluate the effectiveness of \sysname with two commercial VR devices, and the results indicate that it can achieve over $94.2\%$ and $90.5\%$ top-3 accuracy in inferring character-level and word-level keystrokes with varying lengths, respectively.
In addition, empirical results show that \sysname is resilient to several practical impact factors and presents effectiveness in various real-world scenarios, which provides a complementary and orthogonal attack surface for the exploration of keystroke inference attacks in VR platforms.

\end{abstract}


\section{Introduction}
\label{sec:introduction}

Virtual Reality (VR) technology has provided a novel human-computer interaction (HCI) paradigm that revolutionizes the way \major{people} communicate with digital content by creating immersive and interactive virtual environments.
In particular, a VR system usually consists of a headset running particular operating systems and \major{rendering} digital content on the head-mounted display (HMD) as well as two handheld VR controllers to facilitate interactions.
This new mobile platform transcends traditional screen-based displays with spatial interactions alongside multi-sensory feedback, which has led to its immense popularity in recent years.

Despite \major{its} innovative and immersive interactions, \major{people} still have to type keystrokes on a virtual keyboard for certain functionalities, \ie, entering passwords to login accounts, which exposes VR users to a legacy vulnerability existing in common mobile platforms, that is, privacy leakage from keystroke inference.
Recent studies (\eg, \cite{slocum2023going, zhang2023s, wu2023privacy, luo2022holologger, al2021vr, meteriz2022keylogging, su2024remote, wang2024gazeploit, gopal2023hidden, luo2024eavesdropping, ling2019know, khalili2024virtual}) have confirmed and demonstrated the feasibility of keystroke inference attacks in VR \major{devices}. In particular, these attacks target on the VR headset, either by installing malware to obtain motion sensor data~\cite{zhang2023s, slocum2023going, wu2023privacy, luo2022holologger, ling2019know}, recording videos of head movements and hand gestures~\cite{gopal2023hidden, meteriz2022keylogging, khalili2024virtual}, or utilizing side-channel information (\eg, Wi-Fi CSI~\cite{al2021vr}, acoustic signals in pressing controller buttons~\cite{luo2024eavesdropping}, unencrypted packets in network traffic~\cite{su2024remote}, \major{or gaze information of virtual avatars~\cite{wang2024gazeploit}}).
In general, these attacks \major{depend on} \major{building inference} models on various types of data traces to classify keys on a virtual keyboard for keystroke inference.
However, the aforementioned keystroke inference attacks are constrained to specific scenarios and present limited scalability.
Specifically, previous studies (\eg, \cite{wu2023privacy, zhang2023s, slocum2023going, al2021vr, luo2024eavesdropping, meteriz2022keylogging, su2024remote, wang2024gazeploit}) are restricted in closed-world classification for inferring keystrokes and inherently difficult to scale due to their reliance of developing multiple trace-based deep neural networks (DNN).
In addition, several attacks that exploit side-channel information can only be effective under \major{controlled} conditions, \ie, a fixed position without movement~\cite{al2021vr}, a sufficient light condition or silent environment to record videos or audios of hand gestures~\cite{gopal2023hidden, meteriz2022keylogging, luo2024eavesdropping, khalili2024virtual}, \major{or a shared virtual space or specific VR apps to monitor avatars' features~\cite{su2024remote, wang2024gazeploit}}, which also limits their scalability and practicality.
\looseness=-1

\begin{figure}[t]
    \centering
    \includegraphics[width=\linewidth]{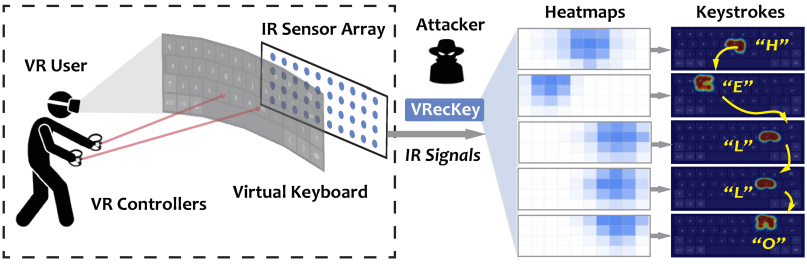}
    \vspace{-0.2in}
    \caption{Illustration of \sysname attack: The victim types the virtual keyboard to input the keystroke ``HELLO''. Meanwhile, the attacker leverages a 2D IR sensor array to capture the IR signals emitted to the ambient environment and reconstruct the heatmaps to infer the virtual keystroke and its trajectory.}
    \vspace{-0.3in}
    \label{fig:motivating_example}
\end{figure}

\begin{table*}[t]
\centering
\scriptsize
\setlength{\tabcolsep}{5pt}
\renewcommand{\arraystretch}{1.2}
\begin{tabular}{cccccccccc}
\hline
\headcol\multicolumn{1}{!{\vrule width 0.5pt}c!{\vrule width 0.5pt}}{\textcolor{white}{\textbf{\begin{tabular}[c]{@{}c@{}}Related VR Attacks\end{tabular}}}} & \multicolumn{1}{c!{\vrule width 0.5pt}}{\textcolor{white}{\textbf{Attack Surface}}} & \multicolumn{1}{c!{\vrule width 0.5pt}}{\textcolor{white}{\textbf{Side Channel}}} & \multicolumn{1}{c!{\vrule width 0.5pt}}{\textcolor{white}{\textbf{NI}}}  & \multicolumn{1}{c!{\vrule width 0.5pt}}{\textcolor{white}{\textbf{NPC}}}  & \multicolumn{1}{c!{\vrule width 0.5pt}}{\textcolor{white}{\textbf{WMI}}} & \multicolumn{1}{c!{\vrule width 0.5pt}}{\textcolor{white}{\textbf{UKI}}} & \multicolumn{1}{c!{\vrule width 0.5pt}}{\textcolor{white}{\textbf{Distance}}} & \multicolumn{1}{c!{\vrule width 0.5pt}}{\textcolor{white}{\textbf{\begin{tabular}[c]{@{}c@{}}Character Level\end{tabular}}}} & \multicolumn{1}{c!{\vrule width 0.5pt}}{\textcolor{white}{\textbf{\begin{tabular}[c]{@{}c@{}}Word Level\end{tabular}}}} \\ \hline
\multicolumn{1}{!{\vrule width 0.5pt}c!{\vrule width 0.5pt}}{TyPose~\cite{slocum2023going}} & \multicolumn{1}{c!{\vrule width 0.5pt}}{Motion sensors in VR headset} & \multicolumn{1}{c!{\vrule width 0.5pt}}{Malware} & \multicolumn{1}{c!{\vrule width 0.5pt}}{\myNocirc} & \multicolumn{1}{c!{\vrule width 0.5pt}}{\myHalfcirc} & \multicolumn{1}{c!{\vrule width 0.5pt}}{\myNocirc} & \multicolumn{1}{c!{\vrule width 0.5pt}}{\myNocirc} & \multicolumn{1}{c!{\vrule width 0.5pt}}{--} & \multicolumn{1}{c!{\vrule width 0.5pt}}{\myNocirc} & \multicolumn{1}{c!{\vrule width 0.5pt}}{\myHalfcirc (82.0\% T-5)} \\ 
\multicolumn{1}{!{\vrule width 0.5pt}c!{\vrule width 0.5pt}}{Zhang \etal~\cite{zhang2023s}} & \multicolumn{1}{c!{\vrule width 0.5pt}}{Motion sensors in VR headset} & \multicolumn{1}{c!{\vrule width 0.5pt}}{Malware} & \multicolumn{1}{c!{\vrule width 0.5pt}}{\myNocirc} & \multicolumn{1}{c!{\vrule width 0.5pt}}{\myHalfcirc} & \multicolumn{1}{c!{\vrule width 0.5pt}}{\myNocirc} & \multicolumn{1}{c!{\vrule width 0.5pt}}{\myNocirc} & \multicolumn{1}{c!{\vrule width 0.5pt}}{--} & \multicolumn{1}{c!{\vrule width 0.5pt}}{\myHalfcirc (93.8\% T-1)} & \multicolumn{1}{c!{\vrule width 0.5pt}}{\myNocirc}  \\ 
\multicolumn{1}{!{\vrule width 0.5pt}c!{\vrule width 0.5pt}}{Wu \etal~\cite{wu2023privacy}} & \multicolumn{1}{c!{\vrule width 0.5pt}}{Motion sensors in VR headset} & \multicolumn{1}{c!{\vrule width 0.5pt}}{Malware} & \multicolumn{1}{c!{\vrule width 0.5pt}}{\myNocirc} & \multicolumn{1}{c!{\vrule width 0.5pt}}{\myHalfcirc} & \multicolumn{1}{c!{\vrule width 0.5pt}}{\myNocirc} & \multicolumn{1}{c!{\vrule width 0.5pt}}{\myNocirc} & \multicolumn{1}{c!{\vrule width 0.5pt}}{--} & \multicolumn{1}{c!{\vrule width 0.5pt}}{\myYescirc (89.7\% T-1)} & \multicolumn{1}{c!{\vrule width 0.5pt}}{\myYescirc (84.9\% T-3)}  \\ 
\multicolumn{1}{!{\vrule width 0.5pt}c!{\vrule width 0.5pt}}{HoloLogger~\cite{luo2022holologger}} & \multicolumn{1}{c!{\vrule width 0.5pt}}{Motion sensors in VR headset} & \multicolumn{1}{c!{\vrule width 0.5pt}}{Malware} & \multicolumn{1}{c!{\vrule width 0.5pt}}{\myNocirc} & \multicolumn{1}{c!{\vrule width 0.5pt}}{\myHalfcirc} & \multicolumn{1}{c!{\vrule width 0.5pt}}{\myYescirc} & \multicolumn{1}{c!{\vrule width 0.5pt}}{\myYescirc} & \multicolumn{1}{c!{\vrule width 0.5pt}}{--} & \multicolumn{1}{c!{\vrule width 0.5pt}}{\myYescirc (73.0\% T-1)} & \multicolumn{1}{c!{\vrule width 0.5pt}}{\myYescirc (89.0\% T-3)}  \\ 
\multicolumn{1}{!{\vrule width 0.5pt}c!{\vrule width 0.5pt}}{VR-Spy~\cite{al2021vr}} & \multicolumn{1}{c!{\vrule width 0.5pt}}{Wi-Fi channel state data} & \multicolumn{1}{c!{\vrule width 0.5pt}}{Wi-Fi CSI data} & \multicolumn{1}{c!{\vrule width 0.5pt}}{\myYescirc} & \multicolumn{1}{c!{\vrule width 0.5pt}}{\myNocirc} & \multicolumn{1}{c!{\vrule width 0.5pt}}{\myNocirc} & \multicolumn{1}{c!{\vrule width 0.5pt}}{\myNocirc} & \multicolumn{1}{c!{\vrule width 0.5pt}}{1.3m} & \multicolumn{1}{c!{\vrule width 0.5pt}}{\myYescirc (69.8\% T-1)} & \multicolumn{1}{c!{\vrule width 0.5pt}}{\myNocirc}  \\ 
\multicolumn{1}{!{\vrule width 0.5pt}c!{\vrule width 0.5pt}}{Meteriz-Y{\i}ld{\i}ran \etal~\cite{meteriz2022keylogging}} & \multicolumn{1}{c!{\vrule width 0.5pt}}{Users' hand gestures} & \multicolumn{1}{c!{\vrule width 0.5pt}}{Hand tracker/Camera} & \multicolumn{1}{c!{\vrule width 0.5pt}}{\myYescirc} & \multicolumn{1}{c!{\vrule width 0.5pt}}{\myHalfcirc} & \multicolumn{1}{c!{\vrule width 0.5pt}}{\myNocirc} & \multicolumn{1}{c!{\vrule width 0.5pt}}{\myNocirc} & \multicolumn{1}{c!{\vrule width 0.5pt}}{0.6--0.8m} & \multicolumn{1}{c!{\vrule width 0.5pt}}{\myYescirc (99.0\% T-1)} & \multicolumn{1}{c!{\vrule width 0.5pt}}{\myYescirc (87.0\% T-5)} \\
\multicolumn{1}{!{\vrule width 0.5pt}c!{\vrule width 0.5pt}}{Su \etal~\cite{su2024remote}} & \multicolumn{1}{c!{\vrule width 0.5pt}}{Unencrypted Photon protocol} & \multicolumn{1}{c!{\vrule width 0.5pt}}{Network traffic} & \multicolumn{1}{c!{\vrule width 0.5pt}}{\myHalfcirc} & \multicolumn{1}{c!{\vrule width 0.5pt}}{\myYescirc} & \multicolumn{1}{c!{\vrule width 0.5pt}}{\myNocirc} & \multicolumn{1}{c!{\vrule width 0.5pt}}{\myYescirc} & \multicolumn{1}{c!{\vrule width 0.5pt}}{--} & \multicolumn{1}{c!{\vrule width 0.5pt}}{\myYescirc (97.6\% T-1)} & \multicolumn{1}{c!{\vrule width 0.5pt}}{\myYescirc (98.1\% T-3)} \\
\multicolumn{1}{!{\vrule width 0.5pt}c!{\vrule width 0.5pt}}{GAZEploit~\cite{wang2024gazeploit}} & \multicolumn{1}{c!{\vrule width 0.5pt}}{Video of users' virtual avatars} & \multicolumn{1}{c!{\vrule width 0.5pt}}{Gaze information} & \multicolumn{1}{c!{\vrule width 0.5pt}}{\myHalfcirc} & \multicolumn{1}{c!{\vrule width 0.5pt}}{\myYescirc} & \multicolumn{1}{c!{\vrule width 0.5pt}}{\myNocirc} & \multicolumn{1}{c!{\vrule width 0.5pt}}{\myYescirc} & \multicolumn{1}{c!{\vrule width 0.5pt}}{--} & \multicolumn{1}{c!{\vrule width 0.5pt}}{\myYescirc (38.7\% T-1)} & \multicolumn{1}{c!{\vrule width 0.5pt}}{\myYescirc (85.9\% T-5)} \\
\multicolumn{1}{!{\vrule width 0.5pt}c!{\vrule width 0.5pt}}{Gopal \etal~\cite{gopal2023hidden}} & \multicolumn{1}{c!{\vrule width 0.5pt}}{Video of VR users' gestures} & \multicolumn{1}{c!{\vrule width 0.5pt}}{Camera} & \multicolumn{1}{c!{\vrule width 0.5pt}}{\myYescirc} & \multicolumn{1}{c!{\vrule width 0.5pt}}{\myHalfcirc} & \multicolumn{1}{c!{\vrule width 0.5pt}}{\myYescirc} & \multicolumn{1}{c!{\vrule width 0.5pt}}{\myNocirc} & \multicolumn{1}{c!{\vrule width 0.5pt}}{3.0--6.0m} & \multicolumn{1}{c!{\vrule width 0.5pt}}{\myYescirc (82.3\% T-1)} & \multicolumn{1}{c!{\vrule width 0.5pt}}{\myYescirc (57.0\% T-3)} \\ 
\multicolumn{1}{!{\vrule width 0.5pt}c!{\vrule width 0.5pt}}{Heimdall~\cite{luo2024eavesdropping}} & \multicolumn{1}{c!{\vrule width 0.5pt}}{Sound from VR controllers} & \multicolumn{1}{c!{\vrule width 0.5pt}}{Acoustic signal} & \multicolumn{1}{c!{\vrule width 0.5pt}}{\myYescirc} & \multicolumn{1}{c!{\vrule width 0.5pt}}{\myHalfcirc} & \multicolumn{1}{c!{\vrule width 0.5pt}}{\myNocirc} & \multicolumn{1}{c!{\vrule width 0.5pt}}{\myNocirc} & \multicolumn{1}{c!{\vrule width 0.5pt}}{1.0--2.2m} & \multicolumn{1}{c!{\vrule width 0.5pt}}{\myYescirc (96.5\% T-1)} & \multicolumn{1}{c!{\vrule width 0.5pt}}{\myYescirc (91.2\% T-5)} \\ 
\hline
\multicolumn{1}{!{\vrule width 0.5pt}c!{\vrule width 0.5pt}}{\textbf{\sysname}} & \multicolumn{1}{c!{\vrule width 0.5pt}}{IR signals from VR controllers} & \multicolumn{1}{c!{\vrule width 0.5pt}}{IR signal} & \multicolumn{1}{c!{\vrule width 0.5pt}}{\myYescirc} & \multicolumn{1}{c!{\vrule width 0.5pt}}{\myHalfcirc} & \multicolumn{1}{c!{\vrule width 0.5pt}}{\myYescirc} & \multicolumn{1}{c!{\vrule width 0.5pt}}{\myYescirc} & \multicolumn{1}{c!{\vrule width 0.5pt}}{2.0--4.0m} & \multicolumn{1}{c!{\vrule width 0.5pt}}{\myYescirc (85.8\% T-1)} & \multicolumn{1}{c!{\vrule width 0.5pt}}{\myYescirc (90.5\% T-3)} \\ \hline
\end{tabular}%
\caption{\major{Comparative analysis with related keystroke inference attacks in VR devices. ``\myYescirc'': Yes, ``\myNocirc'': No, ``\myHalfcirc'': Partly Yes, NI: Non-intrusive, NPC: No physical constraints, WMI: Without model inference, and UKI: Unconstrained keystroke inference.}}
\vspace{-0.2in}
\label{tab:comparison_prior_works_v3}
\end{table*}

Surprisingly, we disclose a novel side-channel attack from the neglected infrared (IR) leakage \major{emitted from} VR controllers, which presents high scalability with fewer \major{physical} constraints than previous keystroke inference attacks.
\major{The causality of IR leakage stems from the fact} that most commercial VR devices rely on the \textit{constellation tracking systems}~\cite{outsideinvsinsideout}
\major{to track} hand movements in typing virtual keystrokes. \major{In this system,} cameras on the headset continuously scan the infrared signals emitted by LEDs embedded in the VR controllers, \major{allowing them to accurately determine the controllers' positions and orientations, as well as the} virtual stick pointing to the virtual keyboard.
As such, due to the line-of-sight (LoS) interactions \major{inherent in} VR platforms, this infrared side channel can be freely captured and extended to \major{other} constellation tracking-based VR systems, \major{which currently dominate} the market of \major{commercial VR devices}.
On the other hand, it is less constrained since it does not \major{require} training models to infer keystrokes \major{indirectly}. Instead, it can directly recognize typed keys from the IR signal trajectory, \major{enabling the reconstruction of} the consecutive keystrokes.
\looseness=-1

\autoref{fig:motivating_example} depicts a typical scenario of our proposed novel side-channel attack.
The victim wears the VR headset and types the word ``HELLO'' on the virtual keyboard, while the attacker exploits the IR signals captured by an IR sensor array consisting of multiple IR sensors to capture IR emissions from the VR controllers.
The captured signals will then be leveraged to reconstruct corresponding keystrokes for \major{uncovering} sensitive user privacy, such as user account passwords.
Though \major{it} appears to be intuitive and straightforward, this new infrared side-channel attack in the VR platform is non-trivial and contains several overlooked challenges as follows.

\begin{packeditemize}
    \item \textbf{\major{Challenge \ding{182}}: Multiple IR Sources.} \major{Most previous studies have overlooked identifying the active typing source, focusing instead on a simplified single-controller typing scenario. However, typing on both controllers can significantly affect signal traces (\eg, motion sensor data~\cite{slocum2023going, zhang2023s, wu2023privacy, luo2022holologger}, acoustic signals~\cite{luo2024eavesdropping}, and network traffic~\cite{su2024remote}). Therefore, identifying which IR source is typing on the virtual keyboard and reducing non-target interference is crucial.}
    \item \textbf{\major{Challenge \ding{183}}: Coordinates Calibration.} In practice, the VR user could type the keyboard floating at different orientations in the virtual scenes, which induces variations in the captured IR signals. Therefore, it is crucial to calibrate the coordinates between the virtual keyboard and the IR receivers by leveraging barely the \major{leaked} IR signals.
    \item \textbf{\major{Challenge \ding{184}}: Scalability Enhancement.} Instead of relying on trace-based DNN models to classify each key for inferring virtual keystrokes~\cite{zhang2023s, slocum2023going, wu2023privacy, meteriz2022keylogging, al2021vr, gopal2023hidden, luo2024eavesdropping, su2024remote, wang2024gazeploit}, it is challenging to enhance the scalability to achieve identifying inputs on the virtual keyboard in a straight and model-free perception manner by monitoring the trajectory of IR leakages.
    
\end{packeditemize}

Given that recent research~\cite{vrcontroller} has revealed that most mainstream VR devices (\eg, Meta Oculus Quest, HTC VIVE, and Sony PlayStation VR) have adopted infrared-based constellation tracking systems, \major{the threat posed by this infrared side channel becomes increasingly significant and concerning.}
\major{Therefore, in this paper, we} propose \sysname, a novel keystroke inference attack to validate the feasibility of the newly identified infrared side channel in VR controllers and understand the leakage of keystrokes on virtual keyboards systematically and comprehensively.
Specifically, we leverage the captured IR signals to generate heatmaps to monitor and \major{mitigate} the image retention to address the issues of multiple IR sources, utilize the IR fluctuation intervals to \major{realize} coordinate calibration, and generate the keystroke trajectory on the keyboard to enhance the scalability and practicality.

In the evaluation, we have assessed the effectiveness of \sysname with a customized IR sensor array on the virtual keyboards of two commodity VR devices, Meta Oculus Quest 2 and PICO 4 All-in-One, while typing character-level and word-level keystrokes with different lengths inside the virtual environments.
Our evaluation results show that \sysname achieves \major{promising} effectiveness in character-level key recognition (T-1 accuracy: $85.8\%$, T-3 accuracy: $94.2\%$) and unconstrained word-level keystroke inference (T-1 accuracy $81.7\%$, T-3 accuracy: $90.5\%$).
Furthermore, \sysname also presents high resilience and transferability when considering several practical impact factors, including different VR devices, orientation angles, cell widths of the customized IR sensor array, attacking distances, typing speeds on the virtual keyboard, and the slight movements of the VR user.
We also investigate the realistic influence of user movements, omnidirectional \major{LED} distributions, and input with single or both controllers.
In addition, we
\major{further evaluate \sysname under three real-world scenarios with varying conditions (\eg, low-visibility environment) to show the practicality of \sysname,}
propose effective countermeasures to defend against this novel side-channel attack, and discuss the ramifications when it comes to the newly-released controller-less VR device, the Apple Vision Pro, as well as illustrate the limitations and future works of \sysname.
\looseness=-1

\autoref{tab:comparison_prior_works_v3} illustrates the properties of \sysname while comparing with \major{ten} state-of-the-art keystroke inference attacks in VR devices (\eg, \cite{slocum2023going, zhang2023s, wu2023privacy, luo2022holologger, meteriz2022keylogging, al2021vr, gopal2023hidden, luo2024eavesdropping, su2024remote, wang2024gazeploit}) \major{qualitatively and quantitatively}, where \sysname presents distinctive advantages of being non-intrusive, model-free, and unconstrained in inferring virtual keystrokes. 
We further summarize our contributions as follows:

\begin{packeditemize}
    \item \textbf{Novel Side-channel Attack Vector.} We introduce a new side-channel attack that leverages the IR signals \major{leaked} from infrared LED lights embedded in VR controllers to infer \major{unconstrained} keystrokes on the virtual keyboard non-intrusively, \major{demonstrating} a new attack vector to understand side-channel vulnerabilities in the constellation tracking system of most commercial VR platforms (\autoref{sec:threat_model}).
    \item \textbf{Customized Attack Design.} We design and implement \sysname, a novel attack to demonstrate the feasibility and scalability of the new infrared side channel, which exploits a customized 2D IR sensor array to capture the leaked IR signals from the two VR controllers, \major{utilizes} the response interval to calibrate the coordinates and reconstruct both character-level and word-level keyboard input inside the virtual environment, as well as the keystroke trajectory without training trace-based DNN models (\autoref{sec:attack_design}). 
    \item \textbf{Comprehensive Evaluations.} We conducted a series of evaluations on two commercial devices, including character-level key recognition and continuous word-level keystroke inference (\autoref{sec:evaluation}). Then, we evaluated \sysname with a set of practical impact factors with consideration of user movements and omnidirectional features of VR controllers \major{in several real-world scenarios} (\autoref{subsec:practical_impact_factors}). The empirical results show that \sysname can effectively reconstruct virtual keystrokes with varying lengths non-intrusively.
\end{packeditemize}

\begin{figure}[t]
    \centering
    \begin{subfigure}[b]{\linewidth}
         \centering
         \includegraphics[width=\linewidth]{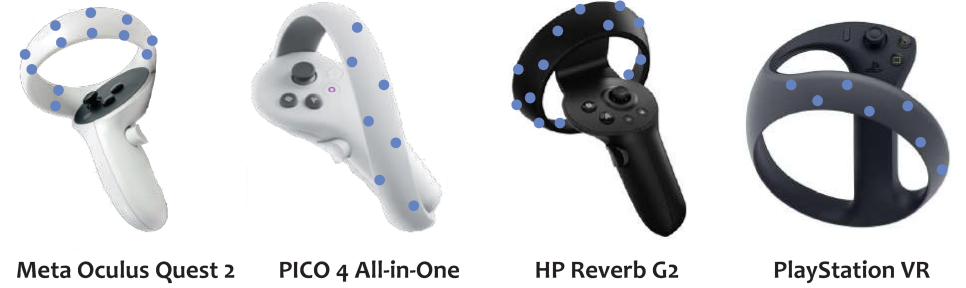}
         \vspace{-0.15in}
         \caption{Infrared LEDs (blue points) on commodity VR controllers.}
         \label{fig:illustration_infrared_led_vr_controller}
    \end{subfigure}
    \begin{subfigure}[b]{\linewidth}
         \centering
         \includegraphics[width=\linewidth]{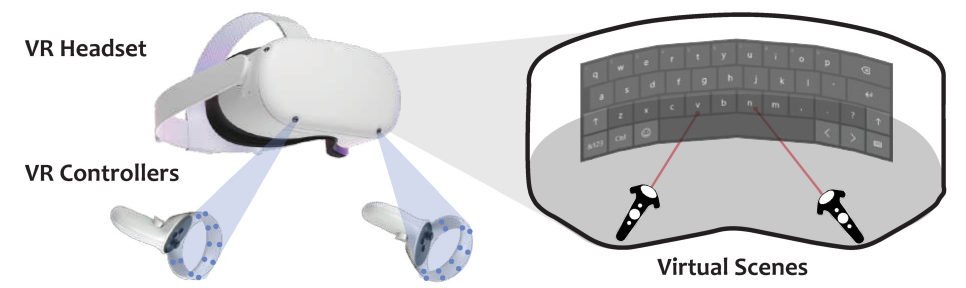}
         \vspace{-0.15in}
         \caption{Constellation tracking system.}
         \label{fig:illustration_constellation_tracking_system}
    \end{subfigure}
    \vspace{-0.15in}
    \caption{Illustration of infrared LEDs embedded in commodity VR controllers (denoted as blue dots on the ring) and the constellation tracking system in VR devices (\autoref{subsec:infrared_led_vr}).}
    \vspace{-0.25in}
    \label{fig:illustration_vr_devices}
\end{figure}

\section{Background}
\label{sec:background}


\subsection{Infrared LED in VR Controllers}
\label{subsec:infrared_led_vr}
Unlike \major{traditional} VR devices such as the HTC VIVE Pro~\cite{htcvivepro}, which rely on external base stations for its outside-in tracking approach, many newly-released VR devices, \ie, the Meta Oculus Quest 2~\cite{oculusquest2}, have \major{utilized} the constellation tracking system~\cite{oculusquest2}, which only consists of a VR headset and two controllers.
Each controller incorporates a set of infrared LEDs discretely positioned on the controller's rings. Simultaneously, the VR headset's cameras continuously capture images of these LEDs' emitted IR signals. 
Then, the constellation tracking system~\cite{gerloni2018immersive} leverages these images to measure the controllers' spatial positions and further infer the user's hand \major{gestures} and body movements.
\autoref{fig:illustration_infrared_led_vr_controller} shows an illustration of four VR controllers of four \major{commercial} VR devices: Meta Oculus Quest 2~\cite{oculusquest2}, PICO 4 All-in-One~\cite{pico4}, HP Reverb G2~\cite{hpreverb}, and PlayStation VR~\cite{ps5nextgenvr}, where we observe that the infrared LEDs are evenly distributed around the controller's ring to track the user's interactions accurately.
\autoref{fig:illustration_constellation_tracking_system} shows the constellation tracking system in these VR devices, 
and such an IR-based constellation tracking system significantly reduces implementation costs as it eliminates the need for the purchase and setup of external base stations~\cite{outsideinvsinsideout}. However, it's important to note that when \major{people} utilize these VR controllers for interaction in virtual scenes, the infrared LEDs on the controllers inevitably emit IR signals into the surrounding environment. These signals may potentially contain sensitive information that could be intercepted and analyzed, posing a potential privacy risk, such as snooping input passwords in virtual \major{scenes}.
\looseness=-1

\begin{figure}[t]
    \begin{subfigure}[b]{.495\linewidth}
         \centering
         \includegraphics[width=\linewidth]{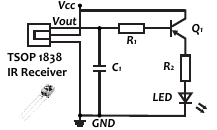}
         \vspace{-0.15in}
         \caption{Circuit of IR sensor.}
         \label{fig:ir_sensor_circuit}
    \end{subfigure}
    \begin{subfigure}[b]{.495\linewidth}
         \centering
         \includegraphics[width=\linewidth]{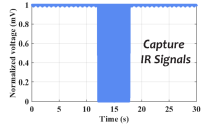}
         \vspace{-0.15in}
         \caption{Recorded voltage signal.}
         \label{fig:ir_sensor_signal}
    \end{subfigure}
    \vspace{-0.2in}
    \caption{Preliminary of IR sensors, including the circuit to capture IR signals and the recorded voltage signal (\autoref{subsec:principle_infrared_sensors}).}
    \vspace{-0.15in}
    \label{fig:ir_sensor_primary}
\end{figure}

\begin{figure*}[t]
    \centering
    \includegraphics[width=\linewidth]{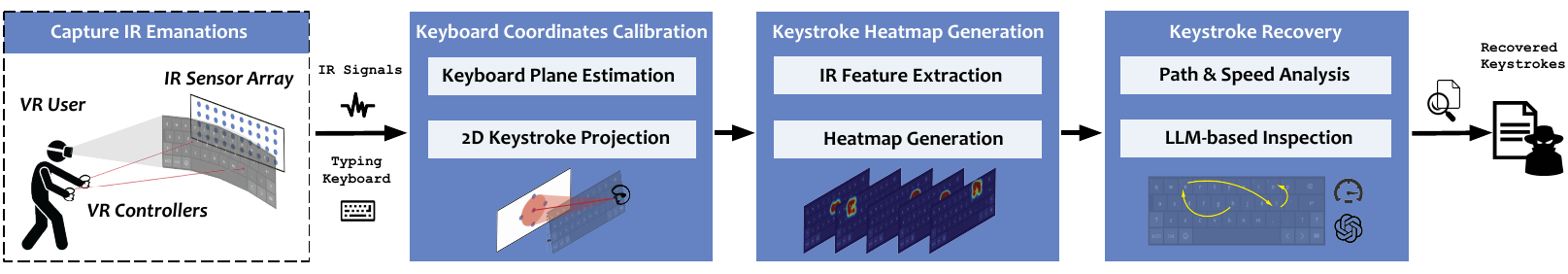}
    \vspace{-0.2in}
    \caption{\major{Overview of \sysname (\autoref{subsec:attack_overview}).}}
    \vspace{-0.2in}
    \label{fig:overview_vrtracker}
\end{figure*}

\subsection{Principle of Infrared (IR) Sensors}
\label{subsec:principle_infrared_sensors}

As discussed above (\autoref{subsec:infrared_led_vr}), the IR signals emitted from VR controllers contain sensitive information, which can be captured by IR sensors (\eg, TSOP 1838 Distance Sensor Receiver~\cite{tsop1838}).
In particular, the embedded photodiode in the IR sensor captures the ambient IR signals emitted from VR controllers, and the built-in circuit demodulates the captured signals to extract the baseband signal from the modulated carrier wave.
Then, the extracted signals are amplified and filtered with a band-pass filter centered around the modulation frequency (\eg, \SI{38}{\kilo\hertz}) to eliminate extraneous noises and enhance signal quality~\cite{daud2013application}.
Then, the IR sensor generates a digital output signal to the connected microcontrollers (MCU), which activate transitions of the output pin to a low state (\SI{0}{\volt}) when receiving an IR signal and reverts to a high state ($V_{cc}$) in the absence of IR signals.
\looseness=-1

\autoref{fig:ir_sensor_circuit} presents a typical IR sensor circuit based on the TSOP 1838 IR receiver, which includes the following components: a resistor $R_{1}$ to pull up the output of TSOP 1838 to $V_{cc}$ when no signal is received, a capacitor $C_{1}$ to filter out noise on the power supply line, a transistor $Q_{1}$ to \major{act} as a switch for turning on the LED when an IR signal is received, a resistor $R_{2}$ to limits the current flowing into the base of $Q_{1}$, and an LED light to depict the reception of IR signals. Assuming the \major{base} current flowing into the transistor $Q_{1}$ is $I_{B}$ and the current gain factor is $h_{FE}$, the output voltage of the IR sensor ($V_{out}$) can be expressed in \autoref{eq:ir_sensor_vout}:
\looseness=-1

\begin{equation}
\label{eq:ir_sensor_vout}
\small
    V_{out} = V_{cc} - (I_{C}\cdot R_{2}) = V{cc} - (h_{FE}\cdot I_{B}\cdot R_{2})
\end{equation}

\noindent When no IR signal is present, the base of the transistor $Q_{1}$ is at a high state, and there are no base current flows ($I_{B}=0$), which makes the output voltage close to $V_{cc}$.

\begin{equation}
\label{eq:ir_sensor_no_ir}
\small
    V_{out} = V_{cc}, I_{B} = I_{C} = 0
\end{equation}

\noindent On the contrary, when an IR signal is received, the output of the TSOP 1838 goes low, which then allows a base current to flow into transistor $Q_{1}$ through the resistor $R_{2}$, which also turns on transistor $Q_{1}$ to allow current to flow from the collector to the emitter side to the LED.
As a results, as shown in \autoref{eq:ir_sensor_ir_present}, the $V_{out}$ is equal to the saturation voltage ($V_{Q(sat)}$) of the transistor $Q_{1}$, \major{which is} typically a small value close to \SI{0}{\volt} as $Q_{1}$ is fully on.

\begin{equation}
\label{eq:ir_sensor_ir_present}
\small
    V_{out} = V_{Q(sat)} \approx 0, V_{cc} \approx I_{C}\cdot R_{2}
\end{equation}

\noindent Therefore, it is feasible to exploit the output voltage ($V_{out}$) of an IR sensor to monitor the presence of IR signals radiated from the VR controllers.
For instance, \autoref{fig:ir_sensor_signal} shows the recorded $V_{out}$ of the IR sensor in capturing IR signals, where we find the corresponding voltage fluctuations when detecting the presence of IR signals emitted from a VR controller of Meta Oculus Quest 2.
Specifically, it records both the spatial (\eg, position) and temporal (\eg, time) information of the VR controller.
Hence, if multiple IR sensors \major{are} at a 2D plane (\eg, IR sensor array), it is feasible to reconstruct the whole trajectory of the VR controller and further \major{infer users' private keystrokes} inside the virtual environment, such as \major{entering passwords in the virtual scenes}.
\looseness=-1

Note that IR signals emitted from VR controllers operate at a specific modulation frequency (\eg, \SI{38}{\kilo\hertz}) that can be captured by the headset's cameras and external IR sensors (\eg, TSOP 1838 IR receivers).
As a result, thermal IR radiations from other objects in the environment (\eg, the human body) cannot interfere with the IR receivers because these IR signals transmit at an unmodulated frequency and present relatively low compared to the strong, focused, and modulated \major{IR} signals from the embedded LEDs on VR controllers.
\looseness=-1

\section{Threat Model}
\label{sec:threat_model}

\begin{figure}[t]
    \begin{subfigure}[b]{.325\linewidth}
         \centering
         \includegraphics[width=\linewidth]{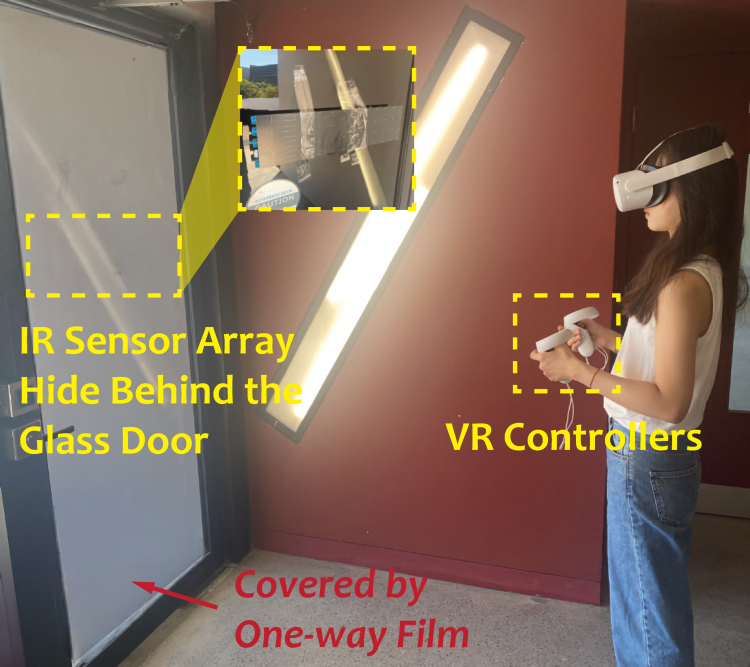}
         \vspace{-0.15in}
         \caption{Concealed.}
         \label{fig:real_world_scenario1}
    \end{subfigure}
    \begin{subfigure}[b]{.325\linewidth}
         \centering
         \includegraphics[width=\linewidth]{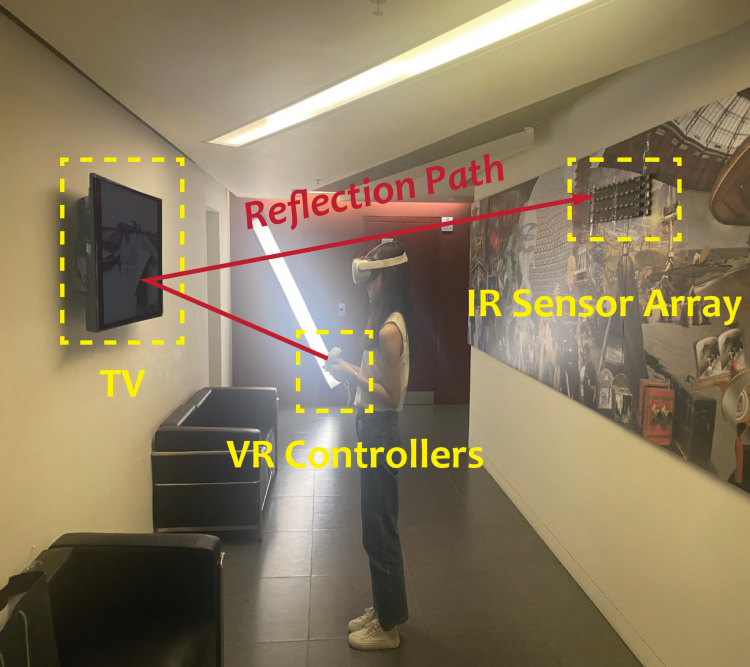}
         \vspace{-0.15in}
         \caption{Reflection-based.}
         \label{fig:real_world_scenario2}
    \end{subfigure}
    \begin{subfigure}[b]{.325\linewidth}
         \centering
         \includegraphics[width=\linewidth]{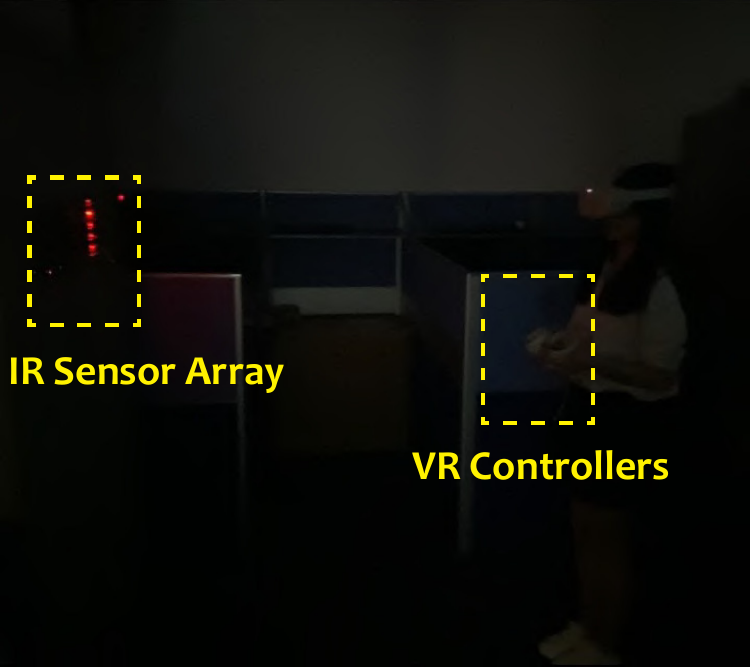}
         \vspace{-0.15in}
         \caption{Low-visibility.}
         \label{fig:real_world_scenario3}
    \end{subfigure}
    \vspace{-0.2in}
    \caption{Three real-world attack scenarios (\autoref{sec:threat_model}), including a concealed attack with one-way film, a reflection-based attack, and an attack in a low-visibility environment.}
    \vspace{-0.2in}
    \label{fig:real_world_scenarios}
\end{figure}

\paragraph{Attack Scenario} We consider a common scenario when the victim wears the VR headset and holds the two controllers to type \major{virtual keystrokes} inside a virtual environment.
Following the research line of deploying attacking devices near VR users (\eg, \major{\cite{gopal2023hidden, al2021vr, meteriz2022keylogging, luo2024eavesdropping, khalili2024virtual}}), we assume that the attacker can place an IR sensor array (\eg, $4\times 10$) consisting of multiple IR sensors \major{near the victim} at a long distance (\eg, \major{\SI{2.0}{\meter}--\SI{4.0}{\meter}}), and remotely analyze the captured IR signals emitted from VR controllers to uncover sensitive keystrokes (\eg, passwords) on the virtual keyboard in an unconstrained manner.
\major{Specifically, the IR sensor array can be stealthily placed either in front of the victim, hidden by a one-way film~\cite{onewayfilm} for concealment, or behind the victim to capture reflected IR signals for keystroke inference, which requires no LoS view to infer virtual keystrokes.}
Such a scenario is prevalent in daily life in various indoor spaces, such as homes and offices, and is \major{plausible for three reasons:} \textit{(i)} most controllers of \major{commercial} VR devices \major{use} infrared-based inside-out tracking systems to realize user interactions with virtual scenes, which inevitably emits IR signals to the \major{surrounding space},
\textit{(ii)} a small IR sensor array positioned or concealed at a considerable distance in front rather than on the VR device's side, is less likely to \major{draw the attention} of the victim \major{immersed} in the virtual \major{environment, especially in low-visibility environments like a dark room},
\major{and \textit{(iii)} a common room has TVs or glass walls, may reflect IR signals, enabling an IR sensor array positioned behind the user to capture these reflections and potentially leak keystroke without LoS view, as demonstrated in relevant studies \cite{huang2023homespy}.}
\major{\autoref{fig:real_world_scenarios} illustrates three real-world attack scenarios: a concealed attack, a reflection-based attack, and an attack in a low-visibility environment, respectively.}
Note that our attack scope primarily targets controller-based VR devices, which constitute the majority of commodity VR products, excluding controller-less VR devices, \ie, Apple Vision Pro~\cite{visionpro}, which is unavailable as of this writing.
\looseness=-1


\paragraph{Attacker's Capability} Unlike previous VR-related attacks for snooping virtual keyboard input (\eg, \cite{wu2023privacy, slocum2023going, zhang2023s, luo2022holologger, ling2019know}), we do not assume the attacker can compromise the VR headset to install malicious software or apps for accessing motion sensor data, nor directly obtain the input keystrokes.
\major{We also do not assume the victim would join unauthorized online meetings to share hands and gaze movements through the virtual avatar~\cite{su2024remote, wang2024gazeploit}.} 
Moreover, we assume the attacker cannot directly monitor the victim's head movements or hand gestures (\eg, placing multiple cameras to record videos~\cite{gopal2023hidden, khalili2024virtual}) \major{as this would easily arouse the victim's suspicion.}
In addition, the victim can type the virtual keyboard displayed in the VR headset while sitting \major{at a} stationary boundary or standing \major{within} the playing zone \major{with natural and casual movements, instead of requiring the victim to sit at constrained} positions between the transceivers in other side-channel attacks~\cite{al2021vr, luo2024eavesdropping}.
\looseness=-1


\section{Attack Design}
\label{sec:attack_design}

\subsection{Overview of \sysname}
\label{subsec:attack_overview}

\autoref{fig:overview_vrtracker} shows the overview of \sysname.
An attacker first places a 2D IR sensor array \major{near} the victim, who wears the VR headset and holds the controllers to type on the virtual keyboard.
When the victim types different keystrokes, the VR controllers emit IR signals to interact with the virtual scenes, which also leaked to be captured by the multiple IR sensors on the array (\autoref{subsec:customized_ir_sensor_array}).
First, \sysname leverages the IR signals from the IR sensors to estimate the orientation between the virtual keyboard and the 2D IR sensor array to calibrate the keyboard coordinates and project the 2D keystrokes (\autoref{subsec:keyboard_coordinates_calibration}).
Then, \sysname extracts time-domain and frequency-domain IR features from the IR signals and generates heatmap overlay onto the virtual keyboard (\autoref{subsec:keystroke_heatmap_generation}).
Finally, the attacker can \major{infer unconstrained} keystrokes by reconstructing the trajectory of the typed \major{keystrokes, estimating the typing speed,} and applying an LLM-based keystroke autocorrection tool to enhance the semantics and grammar of the inferred keystrokes (\autoref{subsec:unconstrained_keystroke_recovery}).

\subsection{\major{Capture IR Emanations and Identify Typing Event}}
\label{subsec:customized_ir_sensor_array}

\paragraph{\major{Customized IR Sensor Array}} To capture the IR signals emitted from the infrared LEDs on the VR controllers, we have designed and implemented a customized 2D IR sensor array
(\autoref{fig:customized_ir_sensor_array} \major{in Appendix}).
It consists of three main components: \textit{(i)} $40$ IR sensors to capture the emitted IR signals and convert them to measurable voltage signals, \textit{(ii)} five Arduino Nano microcontrollers (MCUs) to control the IR sensors, and \textit{(iii)} five microSD card adapters to record the measurable voltage data from the IR sensors.
We have integrated these components together on a custom-built PCB board with a size of \SI{23.5}{}~in $\times$ \SI{6.3}{}~in (\SI{59.7}{\centi\meter}$\times$\SI{16.0}{\centi\meter}).
Specifically, we utilize $40$ KEYES 1838T infrared sensor receiver module boards~\cite{irsensorswebsite} as the IR sensors, and each sensor presents a small size of \SI{1.1}{}~in $\times$ \SI{0.9}{}~in$\times$ \SI{0.3}{}~in (\SI{2.8}{\centi\meter}$\times$ \SI{2.3}{\centi\meter}$\times$ \SI{0.8}{\centi\meter}).
We chose this IR sensor because it can receive the \SI{38}{\kilo\hertz} remote IR signals (\ie, typically \SI{15}{\meter} claimed in the datasheet~\cite{tsop1838tsensor}) through the control of Arduino Nano MCUs and then decode the captured IR signals to be voltage output, and it also shows a promising capability of resisting electromagnetic interference and light.
Furthermore, the default distance between two adjacent IR sensors (\aka, cell width) is set to \SI{5}{\centi\meter} to prevent interference from each other.
Finally, the five Arduino Nano MCUs record all the $40$ voltage output from the $40$ IR sensors at a sampling frequency of \SI{20}{\kilo\hertz}, and then store the data into the five \SI{32}{\giga\byte} microSD cards embedded in the card adapters.
Note that the total cost of building this 2D IR sensor array prototype is approximately $120$ dollars.
\looseness=-1

\begin{figure}[t]
    \begin{subfigure}[b]{.49\linewidth}
         \centering
         \includegraphics[width=\linewidth]{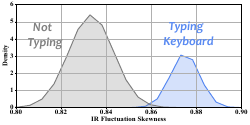}
         \vspace{-0.15in}
         \caption{IR Skewness distribution.}
         \label{fig:typing_event_distribution}
    \end{subfigure}
    \begin{subfigure}[b]{.49\linewidth}
         \centering
         \includegraphics[width=\linewidth]{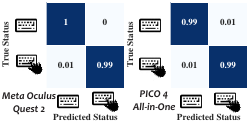}
         \vspace{-0.15in}
         \caption{Identifying Results.}
         \label{fig:typing_event_result}
    \end{subfigure}
    \vspace{-0.05in}
    \caption{Identify typing event by IR signal skewness (\autoref{subsec:customized_ir_sensor_array}).}
    \vspace{-0.2in}
    \label{fig:identify_typing_event}
\end{figure}

\paragraph{\major{Identify Typing Event}} \major{After capturing the leakage of IR emanations, we analyze the signals to detect typing events on the virtual keyboard by focusing on the fluctuation patterns within the IR data.
Specifically, when the captured IR signals contain no typing information, the fluctuations are evenly distributed.
On the contrary, when typing events occur (\eg, button presses on the controller), the IR signals exhibit irregular fluctuations due to the modulation of addresses and commands.
As demonstrated in \autoref{fig:typing_event_distribution}, we calculate the density distribution of skewness in the fluctuating IR signals to distinguish between typing events and static states. \autoref{fig:typing_event_result} further shows that this method achieves $99.3\%$ accuracy in identifying typing events from non-typing statuses on two commercial VR headsets (\ie, Meta Oculus Quest 2 and PICO 4 All-in-One).}
\looseness=-1

\begin{figure*}[t]
    \begin{subfigure}[b]{.43\linewidth}
         \centering
         \includegraphics[width=\linewidth]{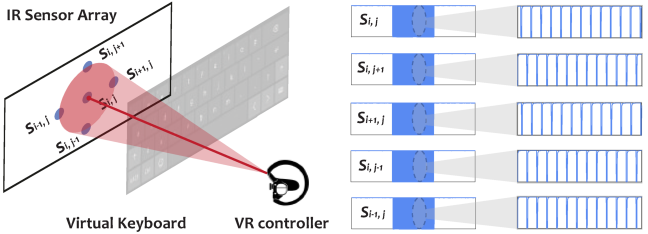}
         \vspace{-0.15in}
         \caption{Virtual keyboard is parallel to 2D IR sensor array.}
         \label{fig:parallel_planes}
    \end{subfigure}
    \begin{subfigure}[b]{.43\linewidth}
         \centering
         \includegraphics[width=\linewidth]{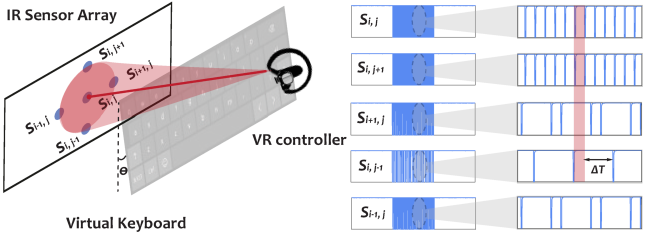}
         \vspace{-0.15in}
         \caption{Virtual keyboard is not parallel to 2D IR sensor array.}
         \label{fig:unparallel_planes}
    \end{subfigure}
    \begin{subfigure}[b]{.12\linewidth}
         \centering
         \includegraphics[width=\linewidth]{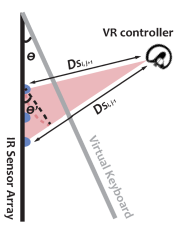}
         \vspace{-0.15in}
         \caption{Illustration.}
         \label{fig:unparallel_planes_side}
    \end{subfigure}
    \caption{Keyboard Coordinates Calibration. We consider whether the virtual keyboard is parallel to the 2D IR sensor array (a) or not (b), and present an illustration of how to measure the orientation angle between the two planes (c) (\autoref{subsec:keyboard_coordinates_calibration}).}
    \vspace{-0.2in}
    \label{fig:keystroke_estimation}
\end{figure*}

\subsection{Keyboard Coordinates Calibration}
\label{subsec:keyboard_coordinates_calibration}

Unlike physical or soft keyboards on other mobile devices (\eg, smartphones, tablets), virtual keyboards \major{are typically displayed in front of the VR user within the virtual environment during a typing process.}
Hence, due to the unparalleled orientation between the keyboard plane and the 2D IR sensor array, the captured IR signals could be distorted.
Therefore, it becomes imperative to conduct keyboard coordinates calibration to calculate the orientation angles and recalibrate the coordinates to align with the keyboard plane. 
Below, we design and implement a two-step process for keyboard coordinates calibration in \sysname, which encompasses keyboard plane estimation and 2D keystroke projection.
\looseness=-1

\paragraph{Keyboard Plane Estimation} \major{As demonstrated in~\cite{meteriz2022keylogging}, virtual keystrokes are aligned on the same plane as virtual keyboards, with the study utilizing a regression model to obtain the keyboard plane based on 3D keystroke data. Drawing inspiration from this work, we propose a model-free method for keyboard plane estimation, leveraging the variations in response times of IR signals captured by multiple sensors to estimate the fitting keyboard plane.}
Assuming the orientation angle between the keyboard plane and the 2D IR sensor array as $\theta$ and the grid width is $l$, we select a specific IR sensor $s_{i, j}$ with four of the adjacent IR sensors (\ie, top: $s_{i, j+1}$, bottom: $s_{i, j-1}$, left: $s_{i-1, j}$, and right: $s_{i+1, j}$) and analyze their captured IR signals to estimate the keyboard plane as shown in \autoref{fig:parallel_planes}.
Specifically, if the two planes are parallel ($\theta = 0^{\circ}$), the four adjacent IR sensors present similar patterns and fluctuations when capturing IR signals as they are evenly distributed on the circumference of the IR emanation frustum.
Meanwhile, due to the closer distance, IR signals captured by $s_{i, j}$ present tight intervals and more drastic fluctuation than the other four sensors.
On the contrary, \autoref{fig:unparallel_planes} shows the condition when the two planes are not parallel ($\theta \neq 0^{\circ}$), and the captured IR signals from $s_{i, j+1}$, $s_{i, j}$, and $s_{i, j-1}$ presenting different intervals.
Specifically, we denote the interval differences between $s_{i, j-1}$ and $s_{i, j+1}$ as $\Delta T$ and the path length are $D_{s_{i, j-1}}$ and $D_{s_{i, j+1}}$, and obtain the length difference $\Delta D$ as follows:
\looseness=-1

\begin{equation}
\label{eq:ke_dist}
\small
     \Delta D = n_{IR}(D_{s_{i, j-1}} - D_{s_{i, j+1}}) \propto 1/\Delta T,
\end{equation}

\noindent where we can measure the distances $D_{s_{i, j-1}}$ and $D_{s_{i, j+1}}$ by moving the VR controller from the IR sensor array's top to the bottom~\cite{benet2002using} to obtain the $\Delta D$ in a time interval of the sensor response frequency $f_{IR}$ (\eg, \SI{38}{\kilo\hertz}), and it contains $n_{IR}$ times of fluctuations in an IR transmission.
Since the distance between the VR controller and the 2D IR sensor array (\eg, \SI{3.0}{\meter}) is usually much larger than the grid width $l$ (\eg, \SI{5.0}{\centi\meter}), the angle $\theta^{'}$ is close to the orientation angle $\theta$, and \autoref{fig:unparallel_planes_side} shows an illustration of deriving the value of $\theta$ as:
\looseness=-1

\begin{equation}
\label{eq:ke_angle}
\small
    \theta \approx \theta^{'} = arcsin(\frac{\Delta D}{2\cdot l\cdot n_{IR}}), n_{IR}\propto f_{IR}.
\end{equation}

\noindent After we obtain the orientation angle $\theta$, we estimate the coordinates of the keyboard plane relative to the 2D IR sensor array plane, and then project the input keystrokes onto the keyboard plane for coordinate calibration.

\begin{figure}[t]
    \centering
    \includegraphics[width=\linewidth]{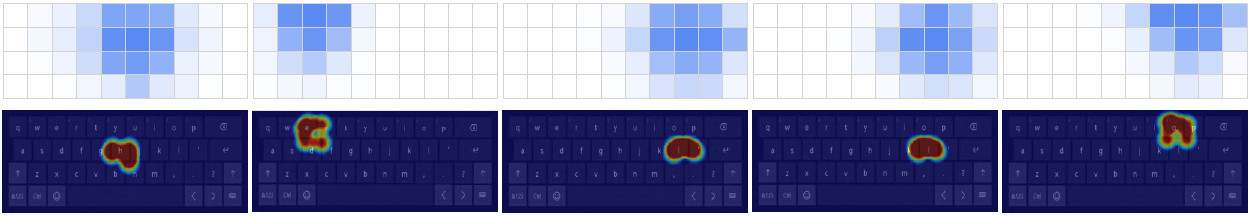}
    \vspace{-0.2in}
    \caption{Keystroke heatmap generation when the VR user types the word \major{``HELLO''} on the virtual keyboard. Upper part: confusion matrices of extracted IR features. Lower part: generated heatmaps on the virtual keyboard (\autoref{subsec:keystroke_heatmap_generation}).}
    \vspace{-0.15in}
    \label{fig:heatmap_cm_example}
\end{figure}

\paragraph{2D Keystroke Projection} Based on the obtained orientation angle $\theta$ between the virtual keyboard and the 2D IR sensor array, we then project the keystrokes detected from the IR sensor array to the virtual keyboard to achieve the calibration of keyboard coordinates.
In practice, we assume the normal vector on the virtual keyboard plane is $N=(A, B, C)$ and the point $P_{0}=(x_{0}, y_{0}, z_{0})$ is on the virtual keyboard, and the virtual keyboard plane can be written as:
\looseness=-1

\begin{equation}
\label{eq:keyboard_plane}
\small
    A(x-x_{0}) + B(y-y_{0}) + C(z-z_{0})=0
\end{equation}

\noindent Furthermore, let $N_{IR} = (A_{IR}, B_{IR}, C_{IR})$ as the normal vector and $P_{IR} = (x_{IR}, y_{IR}, z_{IR})$ be the point of the IR sensor array, the IR sensor array plane is shown \autoref{eq:ir_sensor_array_plane}:

\begin{equation}
\label{eq:ir_sensor_array_plane}
\small
\begin{cases}
    A_{IR}(x-x_{IR}) + B_{IR}(y-y_{IR}) + C_{IR}(z-z_{IR})=0 \\ \\
    cos(\theta) = \frac{N\cdot N_{IR}}{\left\| N \right\|\left\| N_{IR} \right\|} \text{, $\theta$ is the orientation angle.}
\end{cases}
\end{equation}

\noindent Therefore, the projection of point $P_{IR}$ on the virtual keyboard plane is the point $P_{p} = (x_{p}, y_{p}, z_{p})$, where the vector $v=\overrightarrow{P_{IR}P_{0}}$ is the sum of a vector parallel to the virtual keyboard plane and a vector perpendicular to this plane, where $v$ can be expressed in \autoref{eq:vector_pir_p0}:

\begin{equation}
\label{eq:vector_pir_p0}
\small
    v = \overrightarrow{P_{IR}P_{0}} = (x_{0}-x_{IR}, y_{0}-y_{IR}, z_{0}-z_{IR})
\end{equation}

\noindent Finally, we can project the coordinates of the obtained keystrokes from the IR sensor array plane to the virtual keyboard plane by the following \autoref{eq:keystroke_projection}:

\begin{equation}
\label{eq:keystroke_projection}
\small
    \overrightarrow{P_{IR}P_{0}} = \overrightarrow{P_{p}P_{IR}} + G(A, B, C)
\end{equation}

\noindent where $G$ is the scalar from the IR sensor array plane to the virtual keyboard plan. We could substitute these coordinates into the \autoref{eq:ir_sensor_array_plane} and solve the specific value of $G$ and further acquire the projected coordinates $(x_{p}, y_{p}, z_{p})$.
Note that we only consider orientation angles in common conditions ($\theta< 90^{\circ}$) while assuming the pointing vector of VR controllers cannot be paralleled to the keyboard plane.


\subsection{Keystroke Heatmap Generation}
\label{subsec:keystroke_heatmap_generation}

After the calibration of the virtual keyboard coordinates, \sysname then records the captured IR signals from every IR sensor node on the sensor array simultaneously and generates the heatmap for inferring keystrokes.
Specifically, \sysname first creates IR feature maps by extracting time-domain and frequency-domain features from the captured IR signals, and then generates corresponding heatmaps to demonstrate the trajectory of the VR controller,
which can be further exploited for inferring the user's keystrokes inside the virtual environment.
\looseness=-1

\paragraph{IR Feature Extraction} As mentioned above, the converted voltage signal of a \major{commercial} IR sensor fluctuates between \SI{0}{\volt} and $V_{cc}$ (\eg, \SI{5}{\volt}) when capturing the emitted IR signals, which also reflects temporal (\eg, lasting time, fluctuation interval) and spatial (\eg, position) IR-related features.
\major{Specifically, the voltage variance remains below $0.05$ when no IR signals are captured and it exceeds $0.90$ when capturing IR signals.}
Hence, in \sysname, we first apply a moving-variance window with a threshold of $0.1$ to select the informative signal segment from the raw IR signal, as well as the timestamps indicating the starting time and the end time of the IR emanations.
\major{Then, we normalize the amplitude of IR signals to [$0$, $1$] to mitigate the impact of varying strengths of different VR controllers.}
In particular, we extract six time-domain features, including the starting timestamp, duration time, peaks, troughs, mean, and variance.
Furthermore, we then utilize the \textit{Fast Fourier Transform} (FFT) to the IR signal segment and extract the mean of frequency components, the power spectral density, and the entropy.
These features describe \textit{(i)} position on the virtual keyboard the VR controller points at,
\textit{(ii)} duration time when pointing the specific key to represent the potential continuous same keys in many words, \ie, the word \major{``HELLO''} needs to press the key ``L'' twice successively, and \textit{(iii)} the trajectory of the VR controller across the virtual keyboard at varying timestamps.
\looseness=-1

\begin{figure}[t]
    \begin{subfigure}[b]{.495\linewidth}
         \centering
         \includegraphics[width=\linewidth]{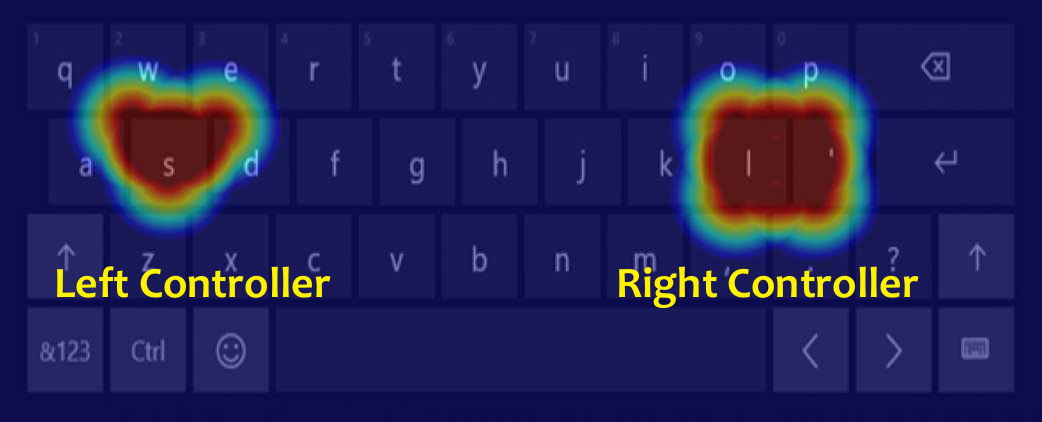}
         \vspace{-0.1in}
         \caption{Heatmap at time $t$.}
         \label{fig:heatmap_before_irr}
    \end{subfigure}
    \begin{subfigure}[b]{.495\linewidth}
         \centering
         \includegraphics[width=\linewidth]{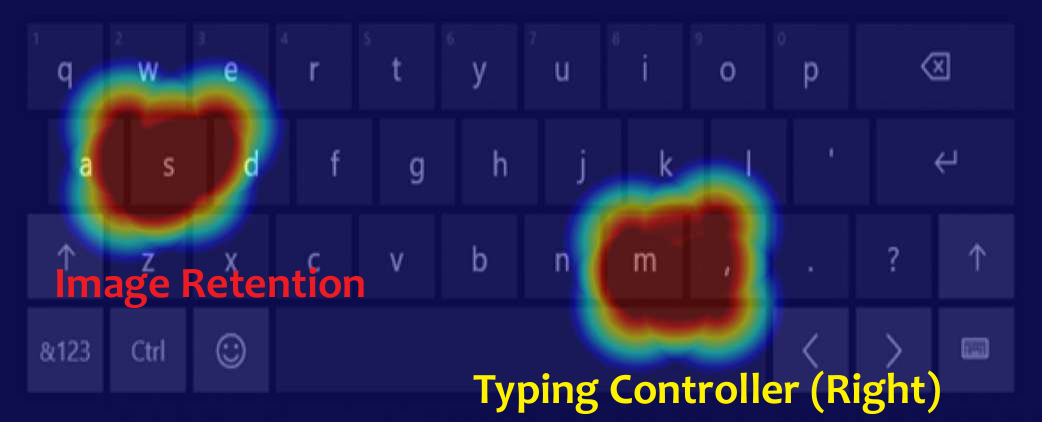}
         \vspace{-0.1in}
         \caption{Heatmap at time $t+\Delta t$.}
         \label{fig:heatmap_after_irr}
    \end{subfigure}
    \vspace{-0.2in}
    \caption{Image retention \major{remove} in a time interval $\Delta t$ (\autoref{subsec:keystroke_heatmap_generation}).}
    \vspace{-0.3in}
    \label{fig:image_retention_removal}
\end{figure}

\paragraph{Heatmap Generation} Subsequently, to visualize the IR features and exhibit the keystroke, \sysname leverages the IR features to generate a heatmap on the IR sensor array to determine the keystrokes.
From \major{the} above discussion, we know that each feature map corresponds to a specific key-typing event on the virtual keyboard. That is, we can map the extracted IR features to the coordinates (\eg, $(x_{i}, y_{i})$) of a specific IR sensor (\eg, $s_{i, j}$) on the IR sensor array.
Let $\mathcal{M}$ represent the mapping function between the IR features and the corresponding keystroke-related heatmap, where $\mathcal{M}$ can be expressed as follows (\autoref{eq:mapping_function}):

\begin{equation}
\label{eq:mapping_function}
\small
    (x_{i}, y_{i}) = \mathcal{M} (\sum{\alpha_{i}f_{t_{i}}} + \sum{\beta_{j}f_{f_{j}}})
\end{equation}


\noindent where $f_{t_{i}}$, $f_{f_{j}}$ represent the $i$th time-domain feature and the $j$th frequency-domain feature, respectively.
The parameters $\alpha_{i}$ and $\beta_{j}$ are the coefficients of the features to balance the weights across diverse dimensions.
Specifically, we leverage Multiple Linear Regression (MLR)~\cite{uyanik2013study} to find coefficients with the closest distance to the IR sensor position $(x_{i}, y_{i})$.
Next, since the IR signals emitted from a VR controller can be captured by other nearby IR sensors, \sysname exploits the output coefficients to ascertain the weights of each IR sensor on the array. These weights normalize them to a range from $0$ to $1$ to facilitate the generation of a heatmap on a $4\times 10$ colored matrix.
In the final stage, \sysname combines the colored matrices generated in
the one-second
time interval of a keyboard-typing (\eg, \major{typically} \SI{0.1}{\second}--\SI{1.5}{\second}~\cite{yang2022eavesdropping}), and utilizes the \textit{applyColorMap} method in Python OpenCV package~\cite{zhao2023odam} to transition the colored matrix into a corresponding heatmap.
Such a heatmap is overlaid on the virtual keyboard and provides a visual representation of keyboard typing events, effectively revealing the trajectory of the keystroke input from the VR controllers.
\looseness=-1

\autoref{fig:heatmap_cm_example} shows an example of confusion matrices of IR features and the generated heatmap from \sysname when the VR user types the word \major{``HELLO''} on the virtual keyboard in the \major{commercial} VR headset, Meta Oculus Quest 2.
Specifically, we set the threshold at $0.8$
\major{to filter out low-value noise while maintaining the spot with the largest probability},
and generate the heatmap with clear spots to reflect the typed keys on the virtual keyboard.
By exploiting these heatmaps, \sysname achieves unconstrained keystroke inference while not relying on training machine-learning-based models for classification, which is widely used in other keystroke inference works (\ie, \cite{al2021vr, luo2022holologger, zhang2023s, slocum2023going, jin2021periscope, cronin2021charger}).
Note that the threshold is determined by the IR signal's strength captured at specific key positions and the key sizes on the generated heatmap.
It can be transferred to different VR devices by mapping the heatmap to the layout of the virtual keyboard.
\looseness=-1

\paragraph{Image Retention \major{Remove}} Despite most people prefer to use only one controller to type on the virtual keyboard, the VR user could enter keystrokes through VR controllers on both left and right hands in a realistic scenario while the IR sensor array may capture IR leakage from multiple \major{IR} sources.
As a consequence, the generated heatmaps could contain image \major{retention} from the VR controller that does not type on the virtual keyboard.
Therefore, we select the two key positions with the highest strength values on the heatmap and monitor their movements in three typing intervals, and then remove the image retentions from the heatmap once we detect the dynamic source, \ie, the typing VR controller.
\autoref{fig:heatmap_before_irr} and \autoref{fig:heatmap_after_irr} individually show the heatmaps of detecting the image retention in a typing interval when the typing VR controller (right hand) moves from key ``L'' to key ``M'', and we can determine the left spot area is the image retention from another untapped VR controller (left hand).
\looseness=-1

\begin{figure}[t]
    \centering
    \includegraphics[width=\linewidth]{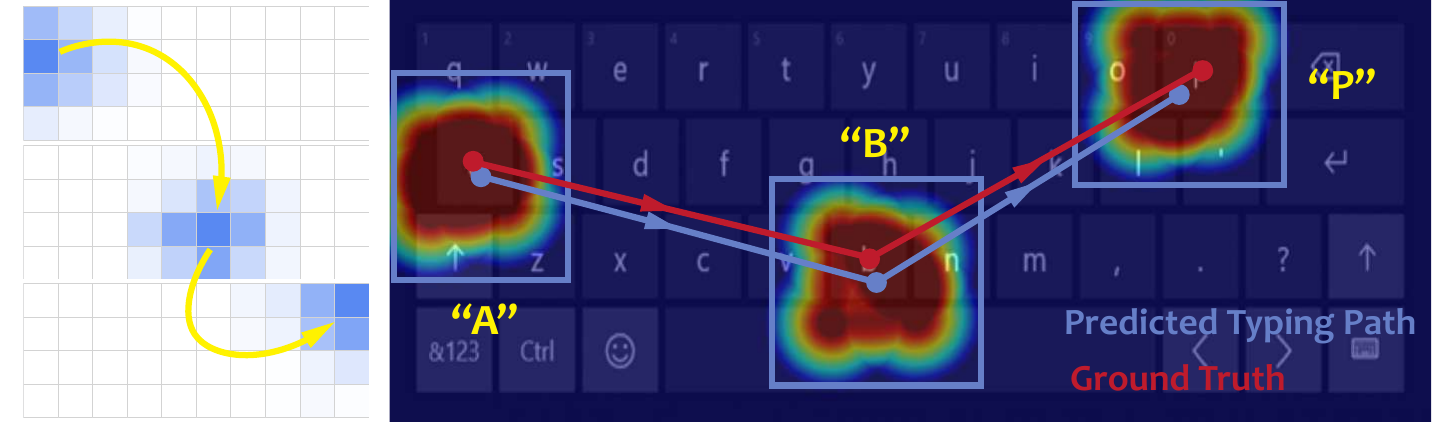}
    \vspace{-0.15in}
    \caption{Example of typing path analysis when typing keys ``A$\rightarrow$B$\rightarrow$P'' continuously on the virtual keyboard (\autoref{subsec:unconstrained_keystroke_recovery}).}
    \vspace{-0.2in}
    \label{fig:example_typing_path_analysis}
\end{figure}

\begin{figure*}[t]
    \minipage{0.34\textwidth}%
        \centering
        \includegraphics[width=.98\linewidth]{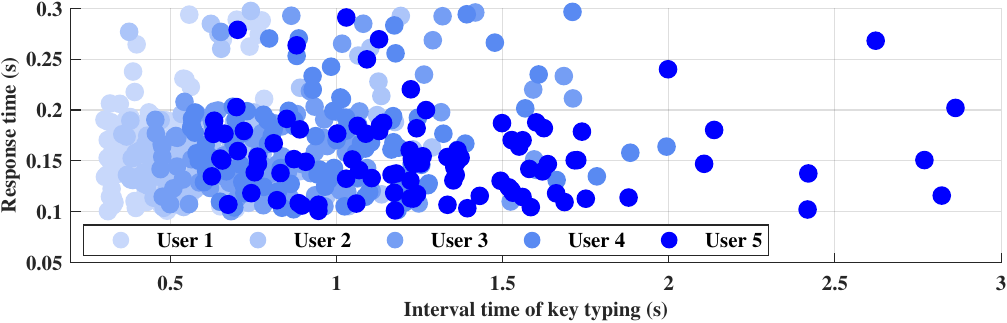}
        \caption{Interval time distribution of five VR users in typing the virtual keyboard.}
        \vspace{-0.2in}
        \label{fig:typing_speed_distribution}
    \endminipage\hfill
    \minipage{0.34\textwidth}
        \centering
        \includegraphics[width=.98\linewidth]{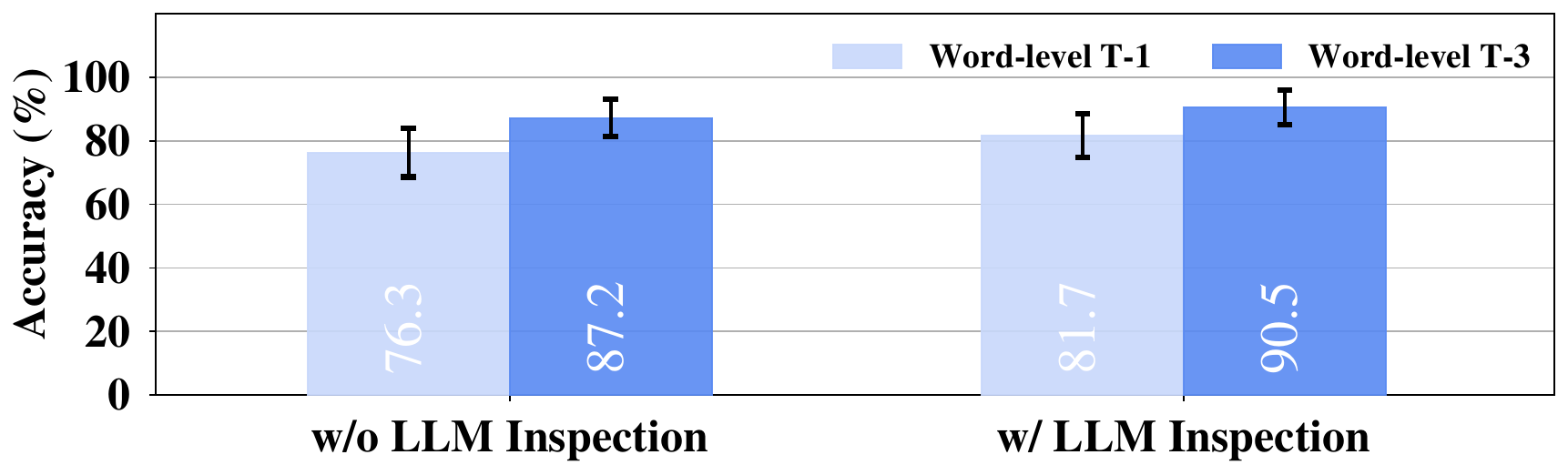}
        \vspace{0.03in}
        \caption{\major{Ablation study results of applying the LLM-based keystroke inspection.}}
        \vspace{-0.2in}
        \label{fig:ablation_study_llm}
    \endminipage\hfill
    \minipage{0.32\textwidth}%
        \centering
        \includegraphics[width=\linewidth]{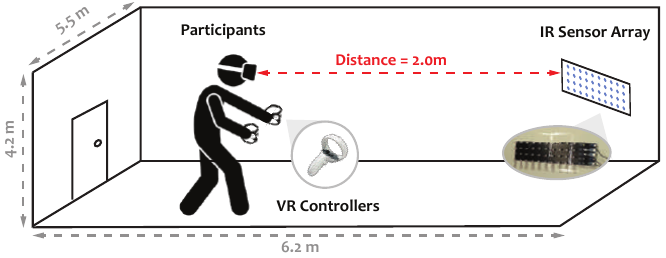}
        \vspace{-0.2in}
        \caption{\major{Default experiment} setup.}
        \vspace{-0.2in}
        \label{fig:experimental_setup}
    \endminipage
\end{figure*}

\subsection{Unconstrained Keystroke Inference}
\label{subsec:unconstrained_keystroke_recovery}

By harnessing the generated heatmaps from the IR signals, \sysname has the ability to identify each individual typing key on the virtual keyboard accurately.
To achieve unconstrained keystroke inference, it is crucial to reconstruct the typing trajectory between consecutive keys. Therefore, \sysname utilizes the bounding box center method to estimate the potential typing trajectory on the virtual keyboard, 
and estimates the VR user's typing speed on the virtual keyboard while distinguishing consecutive identical characters.
Subsequently, it leverages the capabilities of state-of-the-art large language models (LLMs), \ie, ChatGPT, to automatically check the format of \major{reconstructed} keystrokes and \major{address} semantic and grammatical errors.
\looseness=-1

\paragraph{Typing Path Analysis} So far, \sysname has exhibited the ability to recover individual keystrokes \major{by utilizing} captured IR signals.
Typically, a VR user holds the controllers to input keystrokes in a continuous sequence.
This sequence encapsulates not only singular key-pressing events but also the trajectory of the typed keys, revealing the order of the keys in a typed word.
In practice, while typing continuously on the virtual keyboard, the VR controller inevitably sweeps across keys, which triggers an immediate response from the IR sensors along the path that could be falsely recognized as a key-typing event.
Therefore, to mitigate the interference of IR sensors and determine the typing path, \sysname selects the spots on the heatmaps with values higher than the threshold $0.8$ while filtering out interference from low-response \major{($<0.8$)} IR signals \major{(\eg, green and blue zones)}, and then exploits the bounding box center method~\cite{boundingbox} in MATLAB to determine the center from each irregular spot.
Upon acquiring the centers of each bounding box, we connect them together to infer the approximate typing path on the virtual keyboard. For example, \autoref{fig:example_typing_path_analysis} illustrates both the ground truth and \sysname's predictions concerning the typing path as the VR user inputs "A$\rightarrow$B$\rightarrow$P" sequentially, which depicts similar \major{keystroke} trajectories with little deviations.



\paragraph{Typing Speed Estimation} In practice, when it comes to typing words that contain consecutive identical characters, the \major{reconstructed} keystroke trajectory cannot reflect the whole words, \ie, \sysname outputs the same trajectories when typing words like \major{``BEE''} and \major{``BE''}, \major{``OFF''} and \major{``OF''} on the virtual keyboard.
Meanwhile, a recent keystroke study has demonstrated that the typing speed could impact the performance of side-channel keystroke inference~\cite{yang2022eavesdropping}.
Therefore, it is necessary to estimate the typing speed to segment these double-typed characters and reduce its impact on \sysname's performance in unconstrained keystroke inference.
\autoref{fig:typing_speed_distribution} shows the interval time distribution of five different VR users when typing the required keys on the virtual keyboard of Meta Oculus Quest 2 for $100$ times, where we find that the time interval of typing a key ranges approximately from \SI{0.3}{\second} to \SI{2.8}{\second} while the response time of the virtual keyboard ranges from \SI{0.1}{\second} to \SI{0.3}{\second}.
In practice, to estimate the user's typing speed, we define the typing speed at three levels of typing speed: fast (typing interval $<$ \SI{0.5}{\second}), medium (typing interval between \SI{0.5}{\second} and \SI{2.0}{\second}), and slow (typing interval $>$ \SI{2.0}{\second})
\major{, and we collect data samples from all $25$ participants to alleviate user variations.}
As we have set the time interval of heatmap generation as one second (\autoref{subsec:keystroke_heatmap_generation}), if the estimated typing speed falls into the slow level ($>$ \SI{2.0}{\second}), it is possible that the two subsequently generated heatmaps depict the same key on the virtual keyboard.
Once detected, we
concatenate the two same individual keys together and regenerate the output (\eg, \major{``BE''} to \major{``BEE''})
to prevent the double-typing cases that present similar keystroke trajectories, and recover the reasonable keystrokes.
\looseness=-1

\paragraph{LLM-based Keystroke Inspection} Upon obtaining the recovered keys and keystroke trajectory, \sysname generates the outputs of the keystroke typed on the virtual keyboard. 
Given that adjacent keys on the virtual keyboard are in close proximity, misclassified cases exist in recognizing adjacent keys, potentially increasing the incidence of false recognition and impacting \sysname's accuracy.
To mitigate this, we integrate a large language model (LLM) for two purposes:
\textit{(i)} check whether the recovered keystrokes follow a password format,
and \textit{(ii)} perform semantic and grammatical refinement of the output word.
Specifically, we design \major{a zero-shot} LLM-enhanced keystroke inspection tool based on ChatGPT~\cite{wu2023chatgpt} to check keystroke format and correct the predicted words automatically, which is implemented by entering the prompt ``Check if this keystroke \textit{$<$recovered keystrokes$>$} follows a typical password format, otherwise, perform spelling and grammatical check, and then generate top-3 candidates.'' which aims to generate the three most probable alternatives for the keystrokes identified.
\major{\autoref{fig:ablation_study_llm} shows the results of an ablation study in \sysname's performance in word-level keystroke inference when applying the LLM inspection tool or not, where it increases the T-1 and T-3 accuracy by $5.4\%$ and $3.3\%$, respectively.}
Additionally, other online LLMs can also be exploited, \ie, GPT-4~\cite{penteado2023evaluating} and Google Gemini~\cite{saeidnia2023welcome}.
\looseness=-1




\section{Evaluation}
\label{sec:evaluation}

\subsection{Experimental Methodology}
\label{subsec:experimental_methodology}

\paragraph{Experiment Setup} We conduct experiments for the evaluation of \sysname's performance using two commodity VR devices: Meta Oculus Quest 2 and PICO 4 All-in-One, which both adopt the infrared tracking mechanism between the VR headset and the controllers.
In practice, \autoref{fig:experimental_setup}
shows the \major{default experiment setup}, where we ask the \major{participants} to wear the headset and use the controllers to type the required keys.
Specifically, we place the customized IR sensor array (\autoref{subsec:customized_ir_sensor_array}) in front of the VR headset at a primary distance of \SI{2.0}{\meter} and collect the captured IR signals from all IR sensors simultaneously.
\major{In \autoref{subsec:real_world_attack_scenarios}, we conduct further experiments in the three real-world scenarios (\autoref{fig:real_world_scenarios}) with different settings to further demonstrate \sysname's practicality.}
The sampling rate of the five Arduino Nano MCUs is set to \SI{200}{\kilo\hertz}, and the recorded IR sensors' voltage signals are stored in five \SI{32}{\giga\byte} SD cards.
Finally, the collected data samples are processed on a desktop remotely.
\looseness=-1

\paragraph{Participants} We recruited $25$ university students and staffs ($15$ males and $10$ females, with ages ranging from $18$ to $35$) for
the data collection in this study\footnote{\textbf{Ethical Considerations and Open Science Compliance:} We take ethical considerations seriously. This study has obtained \major{the} IRB approval from our institution for data collection, and we only use our accounts of Meta Oculus Quest and PICO platforms to type their default virtual keyboards.
\sysname and our customized IR sensor array have never been released to any other parties. More details (\eg, code, dataset, demo), updates, and appendices will be released on the project website: \url{https://vreckey.github.io/}.}.
Only $11$ participants have prior VR experience, and the other $14$ participants have no knowledge of using VR devices.
We ask the participants to type on the virtual keyboards in VR devices for $30$ minutes to get familiar with VR typing before the official data collection.
All participants were informed that the infrared signals from the VR controllers would be recorded to infer their keystrokes.
During the experiments, participants could move slightly when standing before the IR sensor array \major{while} playing the VR device \major{naturally, which aligns with the experimental settings with most prior studies (\eg,~\cite{slocum2023going, zhang2023s, wu2023privacy, luo2022holologger, al2021vr, meteriz2022keylogging, gopal2023hidden, luo2024eavesdropping, ling2019know}).}
According to our institution's IRB approval, each participant must sign a written consent form that allows us to collect data from human behaviors for evaluation.
\looseness=-1

\paragraph{Evaluation Metrics} To evaluate \sysname's effectiveness, we select \textit{accuracy} as the metric for character-level keystroke recognition, which is defined as the ratio of keystrokes correctly identified as $k$ to the total occurrences of $k$.
In respect of evaluating \sysname's performance in word-level keystroke inference, we select the \textit{top-3 (T-3) accuracy} as the metric since the proposed LLM-based keystroke autocorrection algorithm generates a number of potential candidates after checking the spelling and grammar.
This metric reflects the likelihood of the correct keystroke being present within the first three candidates containing the keystrokes input by the VR user.
\looseness=-1


\subsection{Effectiveness of Character-level Inference}
\label{subsec:effectiveness_single_key}

\paragraph{Data Collection} To evaluate the performance of \sysname in \major{inferring} individual \major{typing} keys, we collect data samples of the IR signals from the 2D sensor array while typing $31$ keys (\ie, $26$ alphabets, space, shift, comma, dot, and enter) through the VR controllers repeatedly.
In practice, each participant presses each key for $100$ times on the default virtual keyboards in the \major{two commercial} VR \major{devices}, Meta Oculus Quest 2 \major{and PICO 4}.
In total, we collect \SI{124000}{} IR signals from the $40$ IR sensors on the 2D IR sensor array when we perform \SI{3100}{} single key-typing actions. 
Furthermore, the collected IR signals can be utilized to generate \SI{3100}{} IR feature maps and corresponding heatmaps overlay onto the virtual keyboard.
Finally, the \major{inferred} keys are compared with the ground truth labels to evaluate \sysname in \major{character-level} key inference.
\looseness=-1

\begin{figure*}[t]
    \centering
    \includegraphics[width=\linewidth]{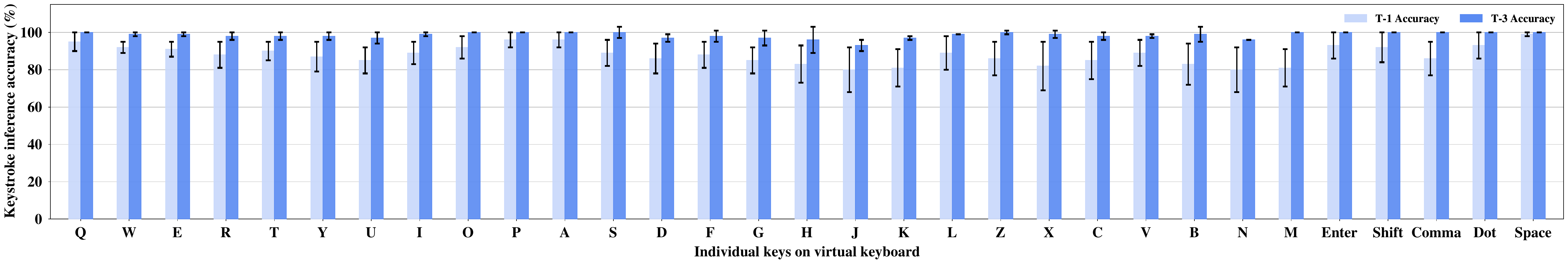}
    \vspace{-0.2in}
    \caption{Evaluation results of \sysname in character-level keystroke inference, including $26$ alphabetic and $5$ special keys.}
    \vspace{-0.15in}
    \label{fig:overall_effectiveness_single_key}
\end{figure*}
\begin{figure*}[t]
    \minipage{0.33\textwidth}%
    \centering
      \includegraphics[width=\linewidth]{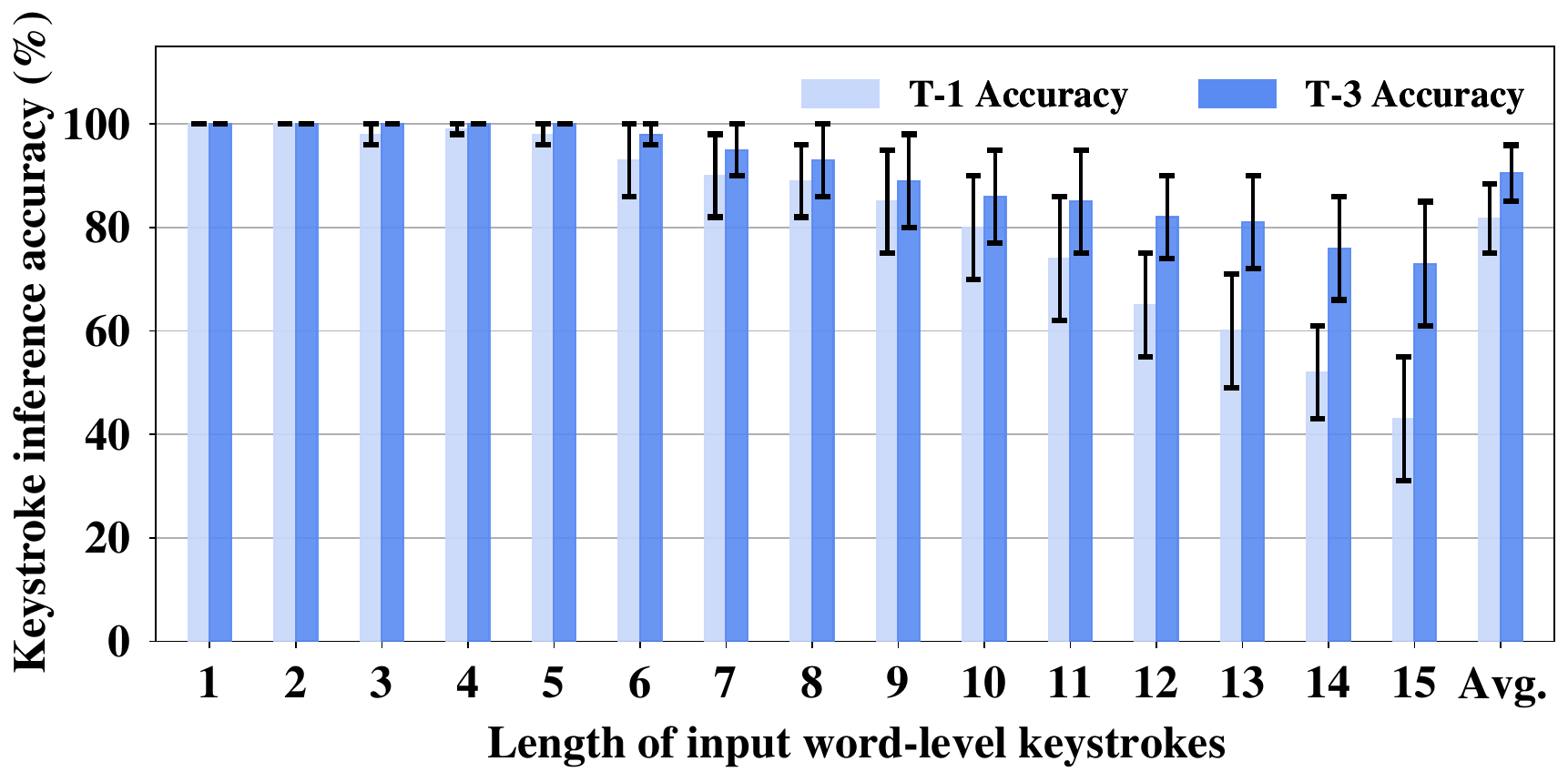}
      \vspace{-0.2in}
      \caption{Evaluation results of \sysname in word-level keystroke inference.}
      \label{fig:overall_effectiveness_keystroke_recovery}
    \endminipage\hfill
    \minipage{0.33\textwidth}%
    \centering
      \includegraphics[width=\linewidth]{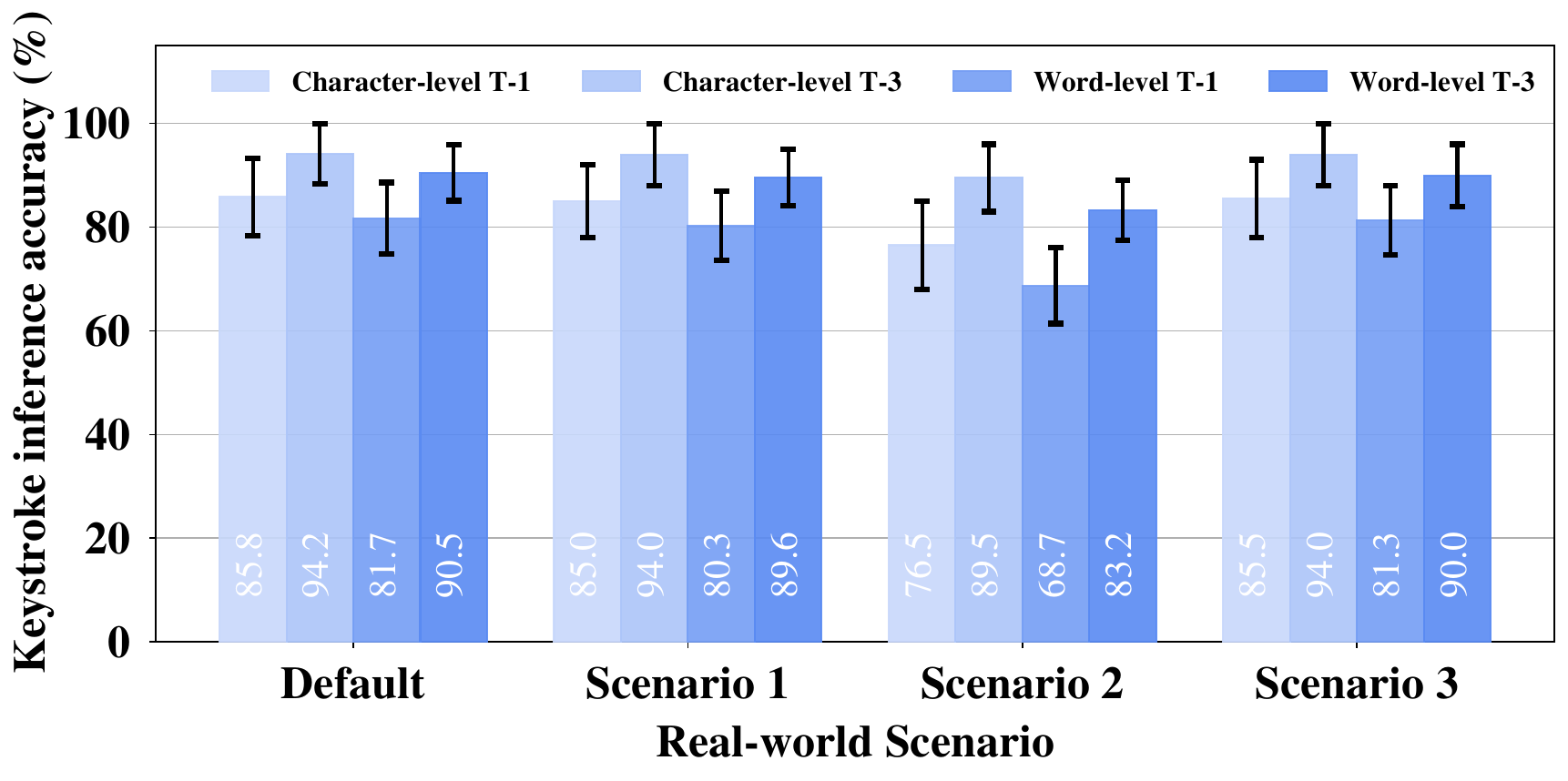}
      \vspace{-0.2in}
      \caption{\major{Evaluation results in the default setting and three real-world scenarios.}}
      \label{fig:overall_effectiveness_real_world_scenarios}
    \endminipage\hfill
    \minipage{0.33\textwidth}
    \centering
      \includegraphics[width=\linewidth]{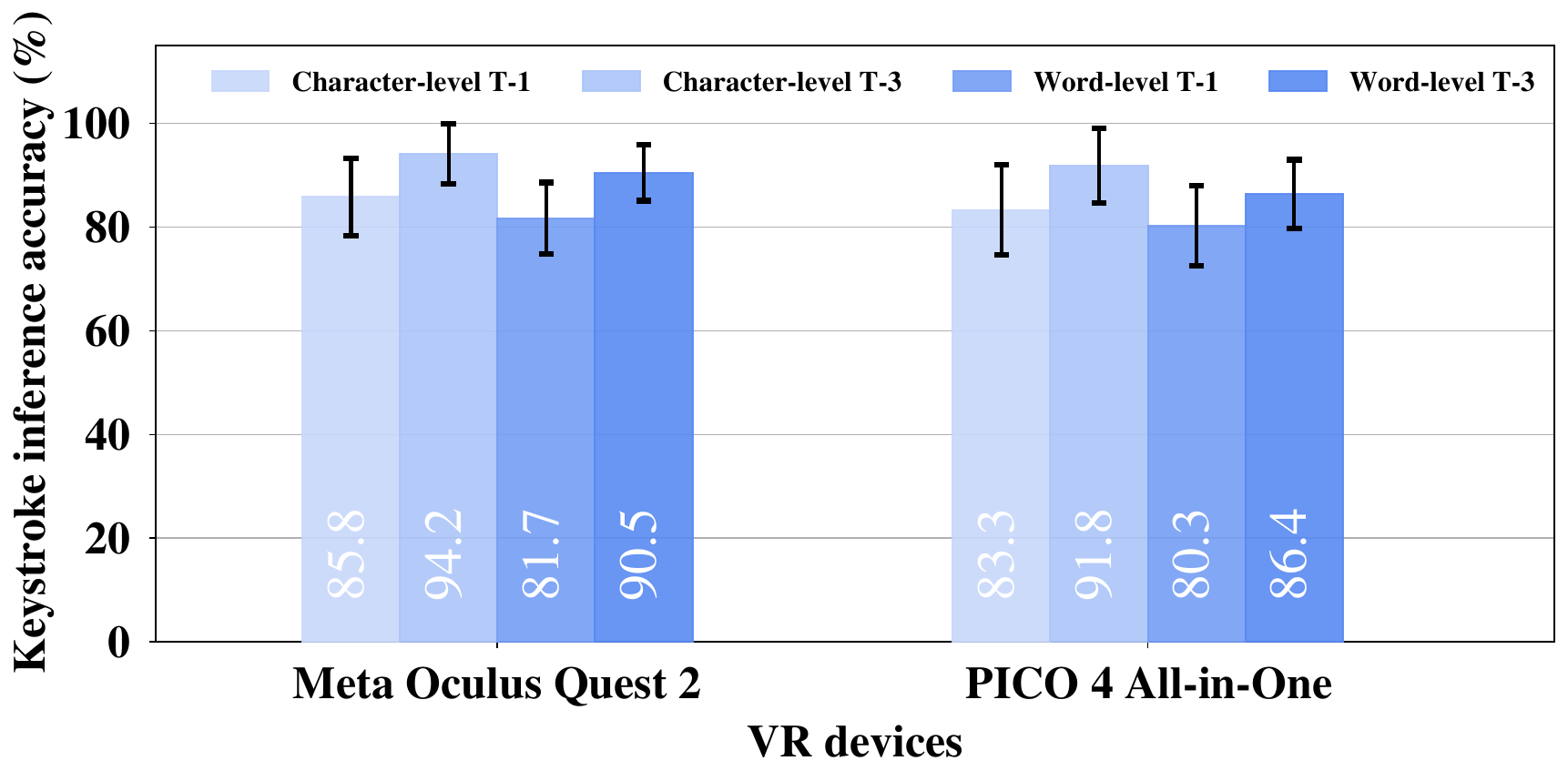}
      \vspace{-0.2in}
      \caption{Evaluation results of two different \major{commercial} controller-based VR devices.}
      \label{fig:if_different_vr_devices}
    \endminipage
    \vspace{-0.1in}
\end{figure*}
\begin{figure*}[t]
    \minipage{0.245\textwidth}%
    \centering
      \includegraphics[width=\linewidth]{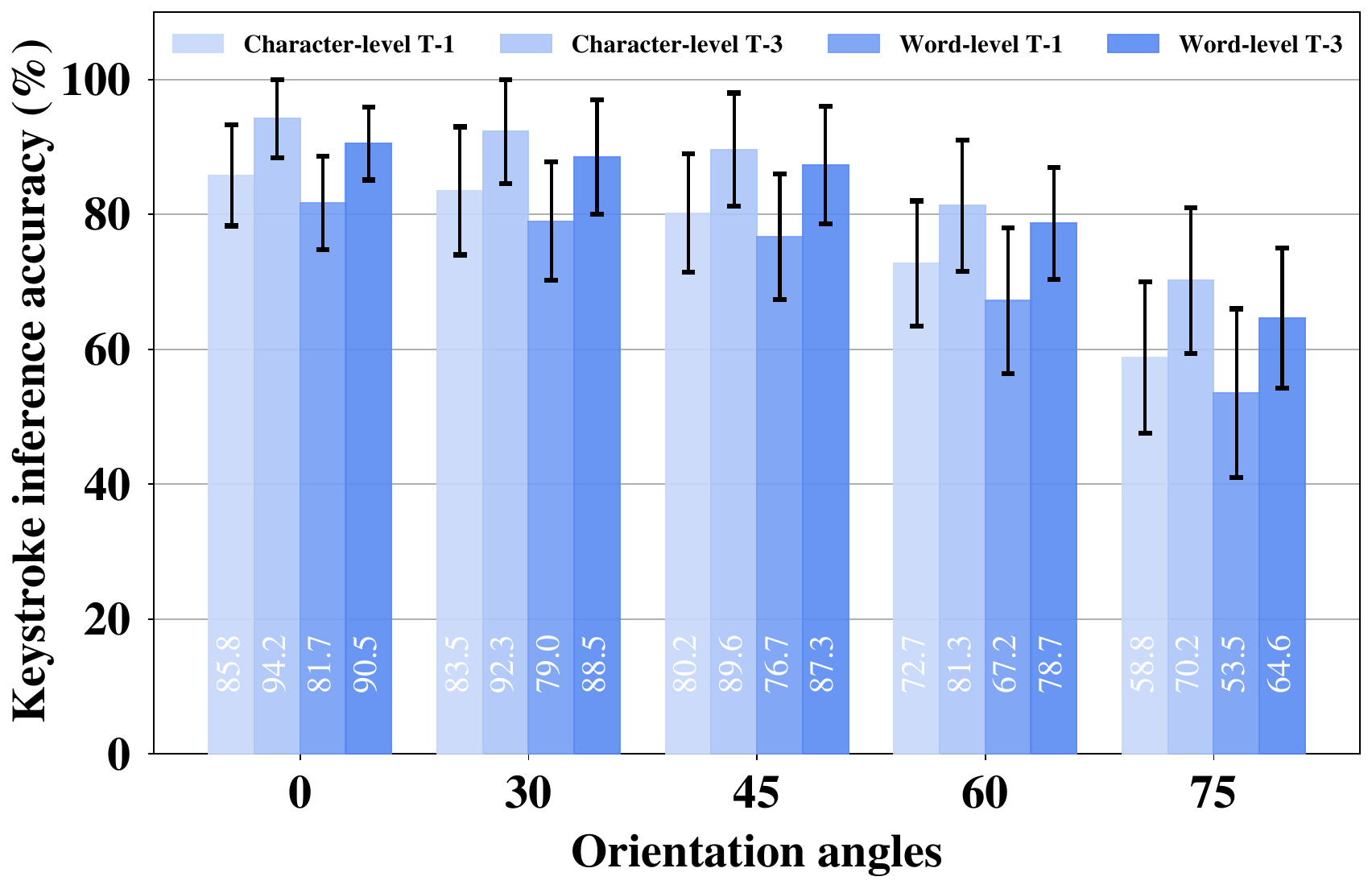}
      \vspace{-0.2in}
      \caption{Evaluation results of three orientation angles.}
      \label{fig:if_different_orientations_positions}
    \endminipage\hfill
    \minipage{0.245\textwidth}%
    \centering
      \includegraphics[width=\linewidth]{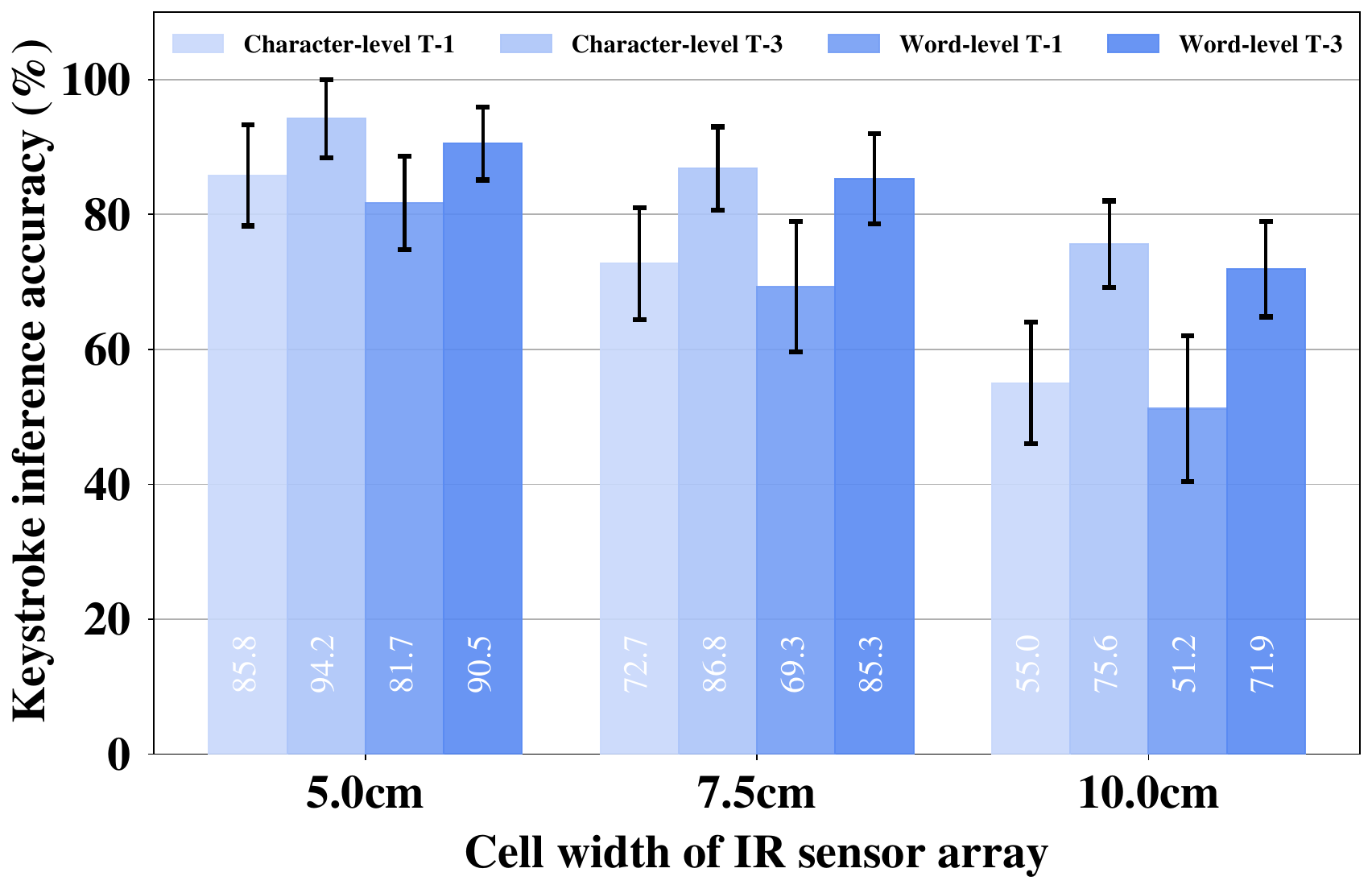}
      \vspace{-0.2in}
      \caption{Evaluation results of three different cell widths.}
      \label{fig:if_different_ir_sensor_array}
    \endminipage\hfill
    \minipage{0.245\textwidth}
    \centering
      \includegraphics[width=\linewidth]{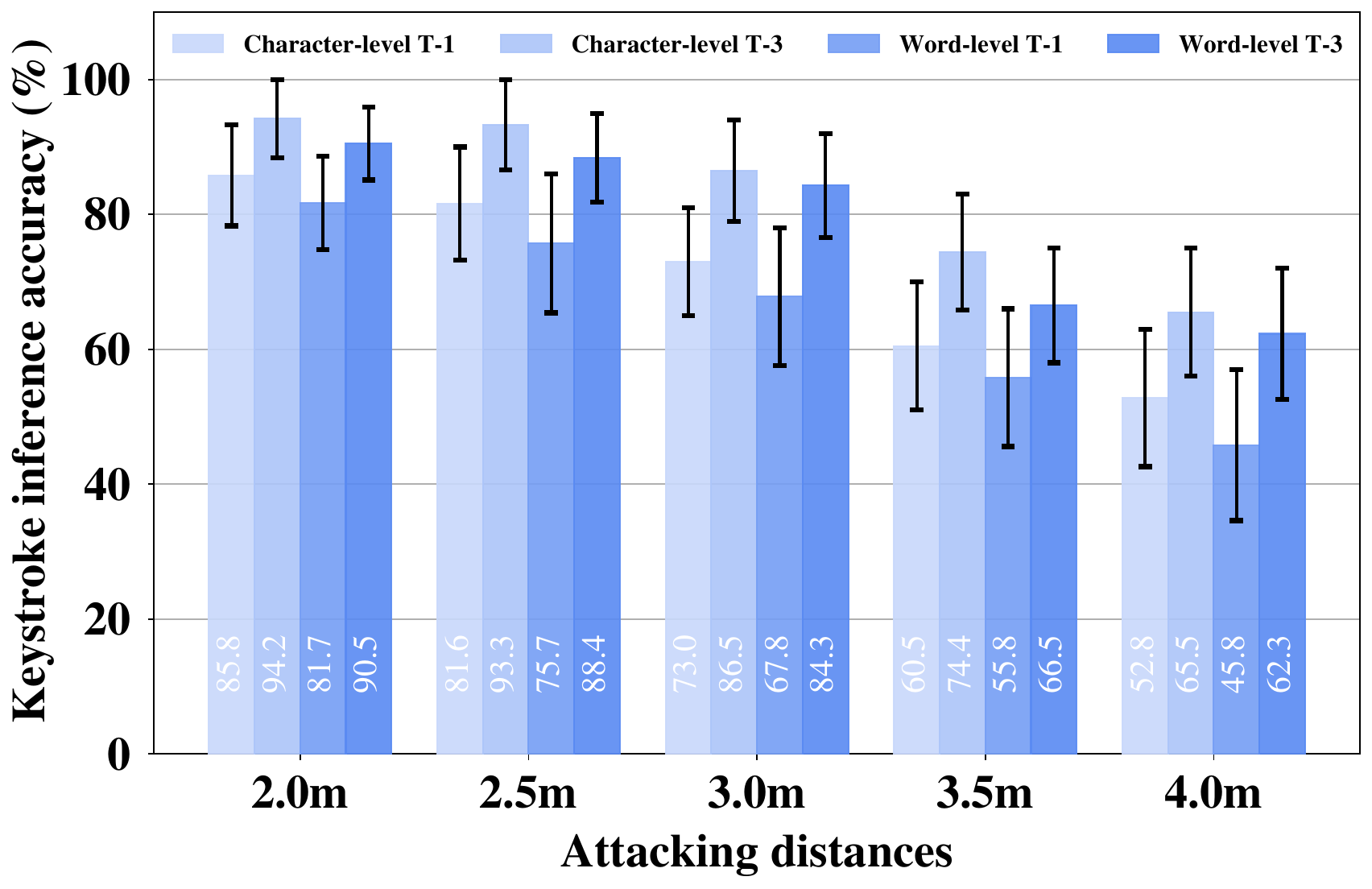}
      \vspace{-0.2in}
      \caption{Evaluation results of five different distances.}
      \label{fig:if_different_attack_distances}
    \endminipage\hfill
    \minipage{0.245\textwidth}
    \centering
      \includegraphics[width=\linewidth]{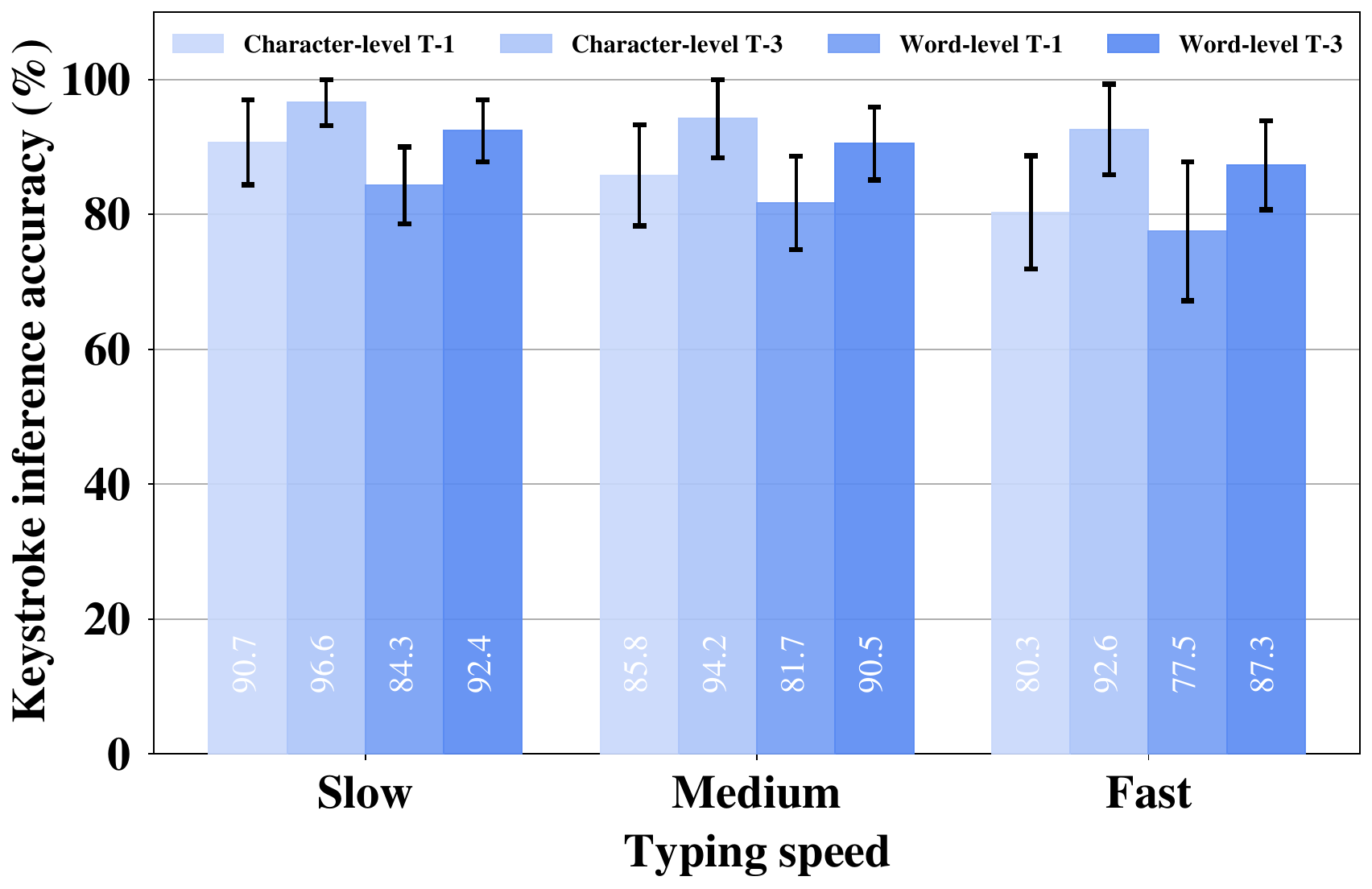}
      \vspace{-0.2in}
      \caption{Evaluation results of three typing speeds.}
      \label{fig:if_different_typing_speeds}
    \endminipage
    \vspace{-0.2in}
\end{figure*}

\paragraph{Character-level Key Inference Results} \autoref{fig:overall_effectiveness_single_key} shows the effectiveness of \sysname in recognizing the $31$ single keys on the virtual keyboard, where it achieves averagely $85.8\%$ T-1 accuracy and $94.2\%$ T-3 accuracy.
In particular, we found that \sysname demonstrates the highest level of accuracy when identifying keys positioned along the border of the virtual keyboard, \ie, alphabetic keys like ``Q'' (T-1: $95\%$, T-3: $100\%$), ``P'' (T-1: $96\%$, T-3: $100\%$), ``A'' (T-1: $96\%$, T-3: $100\%$), and special keys including Enter (T-1: $93\%$, T-3: $100\%$) and Shift (T-1: $92\%$, T-3: $100\%$).
Furthermore, since the Space key occupies a larger layout than normal alphabetic keys, \sysname can recognize it with a higher performance.
On the contrary, the internal keys of the virtual keyboard are more susceptible to being misidentified due to the simultaneous capture of IR signals by adjacent IR sensors. 
That is, a biased typing event on the virtual keyboard could cause these misidentified cases, making \sysname present relatively lower performance, \ie, ``G'' (T-1: $85\%$, T-3: $97\%$), ``H'' (T-1: $83\%$, T-3: $96\%$), and ``J'' (T-1: $80\%$, T-3: $93\%$).
Despite this, \sysname still achieves high T-3 accuracy in both border and internal keys, which depicts its promising performance in the character-level key inference on the virtual keyboard inside the VR headset.
\looseness=-1

\subsection{Effectiveness of Word-level Inference}
\label{subsec:effectiveness_unconstrained_keystroke}

\paragraph{Data Collection} We further evaluate the proposed attack framework, \sysname, in recovering the word-level keystrokes entered by the VR user under more practical attack scenarios (\eg, accounts, passwords).
In practice, we generate alphabetic sequences of high-frequency words ranging in length from one to fifteen, and for each length, we randomly select $100$ words from the Cambridge English vocabulary list~\cite{cambridgeenglish}, respectively.
Then, we ask each participant to type each sequence on the virtual keyboard of the VR headset Meta Oculus Quest 2 and collect corresponding $40$ IR signals from the 2D IR sensor array, where this process is repeated ten times.
In total, for each user, we collect \SI{600000}{} IR signals from the 2D sensor array, extract \SI{15000}{} IR feature maps, generate \SI{15000}{} heatmaps for word-level keystroke inference, and obtain \SI{1500}{} predicted typing path for analysis.
Finally, the recovered keystroke candidates are compared with the ground truth labels to obtain the T-1 and T-3 accuracy to evaluate \sysname in word-level keystroke inference.
\looseness=-1

\paragraph{Word-level Keystroke Inference Results} \autoref{fig:overall_effectiveness_keystroke_recovery} shows the effectiveness of \sysname in recovering word-level keystrokes on the virtual keyboard with lengths ranging from one to $15$, where it achieves an overall T-1 accuracy of $81.7\%$ and T-3 accuracy of $90.5\%$.
Specifically, for keystrokes with lengths less than five, \sysname achieves $99\%$ T-1 accuracy and $100\%$ T-3 accuracy in \major{inferring} these words, whereas its performance degrades drastically when the length of testing keystrokes exceeds ten.
For instance, \sysname achieves only $43\%$ T-1 accuracy and $73\%$ T-3 accuracy when recovering keystrokes with the length of $15$.
To gain a comprehensive understanding of these misidentified cases, our investigation unveiled two primary factors contributing to these occurrences: \textit{(i)} increasing the length of testing keystrokes also leads to more complicated typing paths, which amplifies the possibility of misidentified keys within the keystrokes, and \textit{(ii)} the LLM-based inspection module occasionally generates incorrect words that closely resemble the intended input, especially when the input keystroke deviates from
the correct word.
Nevertheless, it is important to note that \sysname continues to demonstrate competitive performance in word-level keystroke inference under unconstrained conditions when comparing to prior research works \major{(\eg, \cite{slocum2023going, zhang2023s, wu2023privacy, luo2022holologger, al2021vr, meteriz2022keylogging, su2024remote, wang2024gazeploit, gopal2023hidden, luo2024eavesdropping})}.
\looseness=-1

\subsection{\major{Real-world Attack Scenarios}}
\label{subsec:real_world_attack_scenarios}

To validate the stealthiness and practicality of our virtual keystroke inference attack in real-world scenarios, we follow the same procedure and collect data samples from the three settings (\autoref{sec:threat_model}) to evaluate \sysname's end-to-end performance, including a concealed attack scenario, a reflection-based attack scenario, and an attack in a low-visibility scenario.
Specifically, we leverage the Meta Oculus Quest 2 and require all participants to use controllers to type on the default full-size keyboards to enter both character-level and word-level keystrokes in the virtual scene.
\looseness=-1

\paragraph{\major{Scenario \ding{182}: Concealed Attack}} \major{We first conduct the experiments by leveraging commercial one-way film~\cite{onewayfilm} to conceal the IR sensor array behind a \SI{1.1}{\meter}$\times$\SI{2.8}{\meter} floor-to-ceiling window (\autoref{fig:real_world_scenario1}) and ask the participant typing the virtual keyboard at a distance of \SI{2.0}{\meter}. This film allows IR signals to pass through while significantly reducing the visibility of the sensor array, which helps to mitigate the victim’s suspicion. \autoref{fig:overall_effectiveness_real_world_scenarios} demonstrates that the application of the one-way film has minimal impact on \sysname's keystroke inference performance, which maintains competitive accuracy rates for both character-level (T-1: $85.0\%$, T-3: $94.0\%$) and word-level (T-1: $80.3\%$, T-3: $89.6\%$) inference. These results highlight the effectiveness of \sysname in executing stealthy attacks while concealing the IR sensor array, confirming its potential for unobtrusive surveillance.}
\looseness=-1

\paragraph{\major{Scenario \ding{183}: Reflection-based Attack}} \major{We then evaluated \sysname's performance in a reflection-based attack scenario, where the IR sensor array is positioned \SI{1.5}{\meter} behind the VR user standing in front of a reflective surface like a TV or glass wall with proximity of \SI{1.5}{\meter} (\autoref{fig:real_world_scenario2}), and we receive the reflected IR signals to infer keystrokes. This reflection-based setting is reasonable, which aligns with a previous study~\cite{huang2023homespy} and requires only flipping the generated heatmaps vertically to retrieve the correct keystroke trajectory. The results shown in \autoref{fig:overall_effectiveness_real_world_scenarios} indicate a degradation of approximately $4.7\%$--$9.3\%$ and $7.3\%$--$13.0\%$ in character-level (T-1: $76.5\%$, T-3: $89.5\%$) and word-level (T-1: $68.7\%$, T-3: $83.2\%$) keystroke inference, respectively. This decrease is because of the signal attenuation during reflection, which leads to an increase in recognition errors from the heatmaps. Nonetheless, the reflection-based attack demonstrates \sysname's capability in a non-line-of-sight (NLoS) scenario, significantly enhancing the stealthiness of the attack.}
\looseness=-1

\paragraph{\major{Scenario \ding{184}: Low-visibility Attack}} \major{In addition, we also evaluated \sysname in a real-world low-visibility scenario (typically light intensity $<0.1$ lux), such as when users interact with virtual keyboards in a dark room or at midnight (\autoref{fig:real_world_scenario3}) with the default distance settings (\SI{2.0}{\meter}). Under these conditions, traditional camera-based VR keystroke inference attacks (\eg,~\cite{gopal2023hidden, meteriz2022keylogging, khalili2024virtual}) would fail, and enabling the see-through mode in the VR headset would not be feasible. In contrast, our IR sensor array can still capture leaked IR signals to accurately reconstruct virtual keystrokes, achieving high accuracy in both character-level (T-1: $85.5\%$, T-3: $94.0\%$) and word-level (T-1: $81.3\%$, T-3: $90.0\%$) inference, even in very low-visibility conditions. This newly identified infrared side-channel attack significantly extends the threat model of existing keystroke inference attacks on VR platforms, proving highly effective across various scenarios.}
\looseness=-1

\subsection{Practical Impact Factors}
\label{subsec:practical_impact_factors}

\looseness=-1

\paragraph{Different VR Devices} Since different VR devices support various user interfaces, the default virtual keyboards in different \major{commercial} VR devices present alternative layouts.
On the other hand, most of these virtual keyboards adopt a layout similar to full-size QWERTY keyboards that are widely used in other mobile devices (\eg, smartphones and tablets).
These commonalities enhance the potential for extending the applicability of \sysname to target various VR devices.
Thus, to evaluate whether \sysname can launch keystroke inference attacks on different VR devices, we conducted further experiments by separately collecting data samples of IR signals for evaluation from another \major{commercial} VR device, the PICO 4 All-in-One headset.
\autoref{fig:if_different_vr_devices} shows the evaluation results of recovering character-level and word-level keystrokes on the virtual keyboards inside the two VR headsets, where we find \sysname achieves $83.3\%$ T-1 accuracy and $91.8\%$ T-3 accuracy in character-level inference, and $80.3\%$ T-1 accuracy and $86.4\%$ T-3 accuracy in word-level inference on the virtual keyboard of PICO 4 All-in-One, respectively.
In particular, the performance of our proposed attack on the two \major{commercial} VR devices remains consistent, as their default virtual keyboards feature nearly identical layouts.
\looseness=-1

\paragraph{Different Orientations between Virtual Keyboard and IR Sensor Array} In \autoref{subsec:keyboard_coordinates_calibration}, we have proposed methods for estimating orientation angles between the virtual keyboard and the IR sensor array. 
To further explore the impact of different orientation angles, we individually collect data with an orientation angle of $0^{\circ}$ (default settings), $30^{\circ}$, $45^{\circ}$, $60^{\circ}$, and $75^{\circ}$.
\autoref{fig:if_different_orientations_positions} shows the evaluation results of \sysname in keystroke inference at the three mentioned orientation angles. We observe that \sysname decreases to $83.5\%$ T-1 accuracy and $92.3\%$ T-3 accuracy in character-level inference, as well as $79.0\%$ T-1 accuracy and $88.5\%$ T-3 accuracy in word-level inference at the orientation angle of $30^{\circ}$, which shows \major{minimal} performance degradation.
Nevertheless, when adjusting the orientation angle to $75^{\circ}$, character-level inference accuracy decreases to $58.8\%$ T1 and $70.2\%$ T-3 (approximately $25.5\%$ drop), and \major{word-level inference accuracy decreases to} $53.5\%$ T-1 accuracy and $64.6\%$ T-3 accuracy (approximately $27.1\%$ drop).
Hence, the results show the effectiveness of the keystroke coordinates calibration \major{method} in \sysname (\autoref{subsec:keyboard_coordinates_calibration}),
which maintains a \major{promising} keystroke inference performance within \major{common} orientation changes (\eg, less than $60^{\circ}$).
\looseness=-1

\paragraph{Different IR Sensor Arrays with Varying Cell Widths} In the current \major{prototype} of our customized IR sensor array (\autoref{subsec:customized_ir_sensor_array}), we set the cell width between adjacent IR sensors as \SI{5}{\centi\meter}.
As is discussed in \autoref{subsec:keyboard_coordinates_calibration}, multiple adjacent IR sensors could capture the IR signals simultaneously, which also reflect on the generated matrices and heatmaps, \major{which} could \major{affect} the performance of the proposed attack.
To investigate the impact of different cell widths between adjacent IR sensors on \sysname's performance, we have designed and implemented two other IR sensor array boards with cell widths as \SI{7.5}{\centi\meter} and \SI{10.0}{\centi\meter}, respectively.
\autoref{fig:if_different_ir_sensor_array} shows the evaluation results when placing different IR sensor array boards in front of the VR user.
In particular, \sysname achieves $72.7\%$ T-1 accuracy and $86.8\%$ T-3 accuracy in character-level inference, as well as $69.3\%$ T-1 accuracy and $85.3\%$ T-3 accuracy in word-level inference when applying the IR sensor array with \SI{7.5}{\centi\meter} cell width.
In addition, when we select the IR sensor array with a cell width of \SI{10.0}{\centi\meter}, the performance of \sysname decreases to $55.0\%$ T-1 accuracy and $51.2\%$ T-3 accuracy in recognizing character-level keys, and $75.6\%$ T-1 accuracy and $71.9\%$ T-3 accuracy in recovering \major{unconstrained} word-level keystrokes.
The findings illustrate that enlarging the cell width between adjacent IR sensors can result in a larger positional bias in \major{capturing} IR signals, 
\major{which} significantly \major{degrades} the performance \major{and stealthiness} of \sysname.
\major{Therefore, we selected the fine-tuned configuration of \SI{5}{\centi\meter} cell width in designing the 2D IR sensor array.}
\looseness=-1

\paragraph{Different Attacking Distances between VR Controllers and IR Sensor Array} In our primary experiments in \autoref{fig:experimental_setup}, we set the distance between the VR controllers and the 2D IR sensor array as \SI{2.0}{\meter}. Nevertheless, the varying attacking distances could affect the captured IR sensors because of the signal attenuation and interference from the surrounding environment~\cite{huang2023homespy}.
Hence, to investigate the impact of different attacking distances on the performance of \sysname, we place the 2D IR sensor array at different attacking distances: \SI{2.0}{\meter}, \SI{2.5}{\meter}, 
\SI{3.0}{\meter}, \SI{3.5}{\meter}, and \SI{4.0}{\meter}, and collect data samples of IR signals when typing on the virtual keyboard at each attacking distance to evaluate \sysname, respectively.
\autoref{fig:if_different_attack_distances} shows the evaluation results at different attacking distances.
When the attacking distance is set to \SI{2.5}{\meter}, \sysname achieves $81.6\%$ T-1 and $93.3\%$ T-3 accuracy in character-level inference, and $75.7\%$ T-1 and $88.4\%$ T-3 accuracy in word-level inference.
Furthermore, when the attacking distance is \SI{4.0}{\meter}, \sysname exhibits the performance of $52.8\%$ T-1 and $45.8\%$ T-3 accuracy in character-level inference, and $65.5\%$ T-1 and $62.3\%$ T-3 accuracy in word-level inference, respectively.

Hence, we notice that \sysname's performance decreases with the increasing of the attacking distance between the VR controllers and the IR sensor array because of the attenuation of emitted IR signals from the infrared LED and the interference from surrounding environments.
Moreover, we found that the IR sensor array is unable to capture IR signals when the attacking distance is over \SI{5.0}{\meter}, which is much shorter than the typical receiving range of the KEYES 1838T infrared sensor receiver module boards (\ie, typically \SI{15}{\meter}~\cite{irsensorswebsite}).
A reasonable explanation could be the limited strength of IR signals emitted from the infrared LEDs on the VR controllers, which are designed for short-range communications between the cameras on the VR headset and the infrared LEDs on the controllers.
For instance, most commercial-off-the-shelf (COTS) infrared LEDs used by commodity VR controllers present radiant intensity ranging from \SI{4}{\milli\watt} to \SI{125}{\milli\watt} (\eg, OSRAM SFH 4055~\cite{sfh4055}: $4$--\SI{12.5}{\milli\watt}, Vishay TSAL6400~\cite{tsal6400}: $25$--\SI{125}{\milli\watt}, and Everlight IR333-A~\cite{ir333a}: $7.8$--\SI{20}{\milli\watt}), while resulting the insensitivity of being captured by the IR sensors.
Nevertheless, our experiments still show that \sysname realizes unconstrained keystroke inference with \major{acceptable} accuracy at practical attacking distances.
\looseness=-1


\paragraph{Different Typing \major{Speeds}} Previous studies have demonstrated that the typing speed on soft \major{keyboards} could influence the performance of keystroke inference attacks~\cite{yang2022eavesdropping, jin2021periscope, hu2023password}.
In \autoref{subsec:unconstrained_keystroke_recovery}, we have demonstrated that a VR user usually types the virtual keyboard at a time interval ranging from \SI{0.3}{\second} to \SI{2.8}{\second}
, and different typing speed leads to different response times of the captured IR signals, which may impact the \sysname's performance in keystroke inference.
Therefore, to understand the impact of VR controllers' typing speed on the virtual keyboard, we separately collect data samples of IR signals by typing the keyboards at three levels of typing speed: fast (typing interval $<$ \SI{0.5}{\second}), medium (typing interval between \SI{0.5}{\second} and \SI{2.0}{\second}), and slow (typing interval $>$ \SI{2.0}{\second}).
\autoref{fig:if_different_typing_speeds} shows the evaluation results when we \major{individually} type the virtual keyboard inside the Meta Oculus Quest 2 at a faster speed and a slower speed than the primary experiments, where we type virtual \major{keystrokes} at medium speed.
The results demonstrate that \sysname achieves $90.7\%$ T-1 accuracy and $96.6\%$ T-3 accuracy in character-level inference, and $84.3\%$ T-1 accuracy and $92.4\%$ T-3 accuracy in word-level inference when typing on the virtual keyboard at the slow speed.
By contrast, when typing on the virtual keyboard at a faster speed level, \sysname only presents the performance of $80.3\%$ T-1 accuracy and $92.6\%$ T-3 accuracy in character-level inference, and $77.5\%$ T-1 accuracy and $87.3\%$ T-3 accuracy in word-level inference.
We have determined that there is a notable performance degradation of approximately $10.4\%$ and $6.8\%$ T-1 accuracy rates in \major{character-level} and \major{word-level} inference, respectively, when comparing the slow speed level to the fast speed level. This is attributed to the fact that a rapid typing speed results in shorter time intervals focused on the keys of the virtual keyboard, leading to less effective duration features for accurate keystroke inference.
\looseness=-1


\begin{figure}[t]
    \begin{subfigure}[b]{.49\linewidth}
         \centering
         \includegraphics[width=\linewidth]{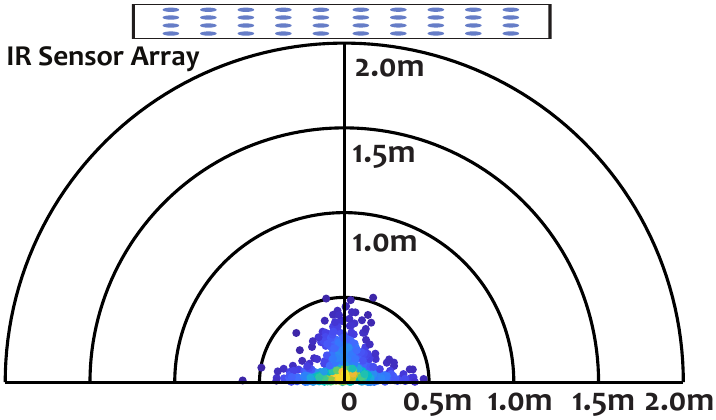}
         \vspace{-0.15in}
         \caption{User movement distribution.}
         \label{fig:user_movement_distribution}
    \end{subfigure}
    \begin{subfigure}[b]{.49\linewidth}
         \centering
         \includegraphics[width=\linewidth]{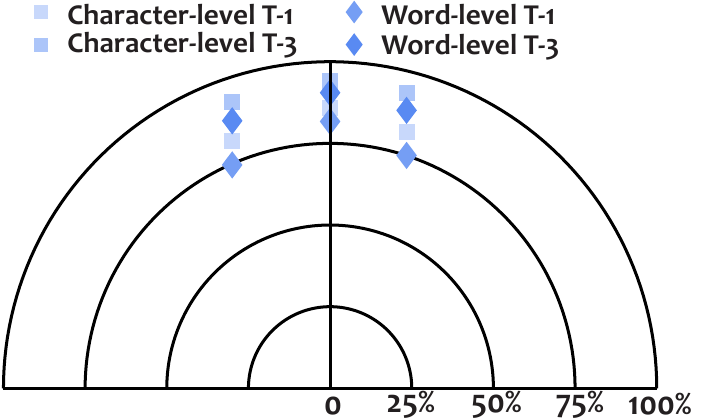}
         \vspace{-0.15in}
         \caption{Keystroke inference results.}
         \label{fig:user_movement_results}
    \end{subfigure}
    \vspace{-0.05in}
    \caption{User movement analysis, including VR users' movements and keystroke inference results (\autoref{subsec:user_movement_analysis}).}
    \vspace{-0.2in}
    \label{fig:user_movement}
\end{figure}

\subsection{User Movement Analysis}
\label{subsec:user_movement_analysis}

In \autoref{subsec:experimental_methodology}, we have illustrated that we allow participants to move \major{casually and naturally} when standing before the IR sensor array as they \major{used} the VR controllers to type on the virtual keyboard.
Nevertheless, VR users' movements during the process of typing on the virtual keyboard could impact \sysname's performance in keystroke inference.
To further understand the user's movements and the potential influence, we record the VR user's standing point distribution in front of the IR sensor array during the typing process, as shown in \autoref{fig:user_movement_distribution}.
It shows that the VR user's movements are concentrated in the range between approximately \SI{0}{\meter} and \SI{0.6}{\meter} in both horizontal and vertical axes.
We evaluated the keystroke inference performance of \sysname when the VR user moves to the leftmost (\SI{-0.60}{\meter}) and rightmost (\SI{0.47}{\meter}) points.
\autoref{fig:user_movement_results} shows character-level and word-level keystroke inference results when launching \sysname under the user's different moving statuses.
In particular, \sysname presents the highest performance when standing exactly in front of the IR sensor array.
When the user moves to the leftmost point, the performance decreases \major{by} $4.5\%$ T-1 and $1.6\%$ T-3 accuracy in character-level inference, and $5.0\%$ T-1 and $3.2\%$ T-3 accuracy in word-level inference.
Meanwhile, when moving to the rightmost, the character-level inference accuracy decreases by $3.8\%$ T-1 and $0.9\%$ T-3, and word-level inference \major{accuracy} decreases by $6.7\%$ T-1 and $2.2\%$ T-3.
Therefore, the results of user movement analysis have demonstrated that the VR user's movements during the typing process lead to limited impact on the keystroke inference from \sysname, especially presenting a neglectable impact on T-3 accuracy.
\looseness=-1

\begin{figure}[t]
    \begin{subfigure}[b]{.495\linewidth}
         \centering
         \includegraphics[width=\linewidth]{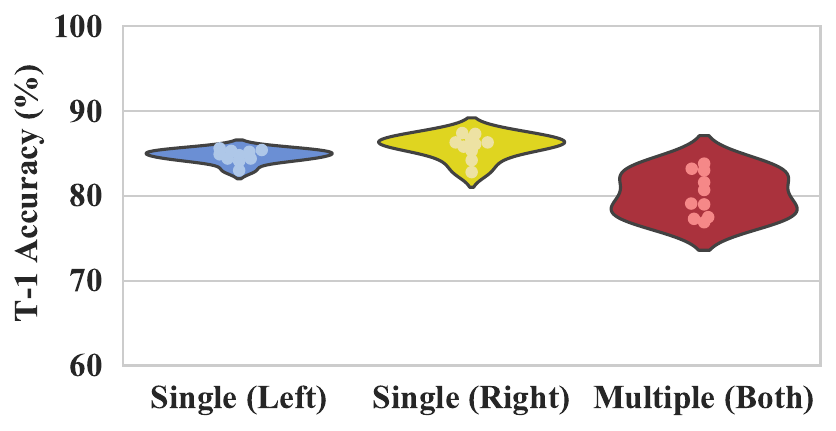}
         \caption{Character-level inference.}
         \label{fig:ir_source_char}
    \end{subfigure}
    \begin{subfigure}[b]{.495\linewidth}
         \centering
         \includegraphics[width=\linewidth]{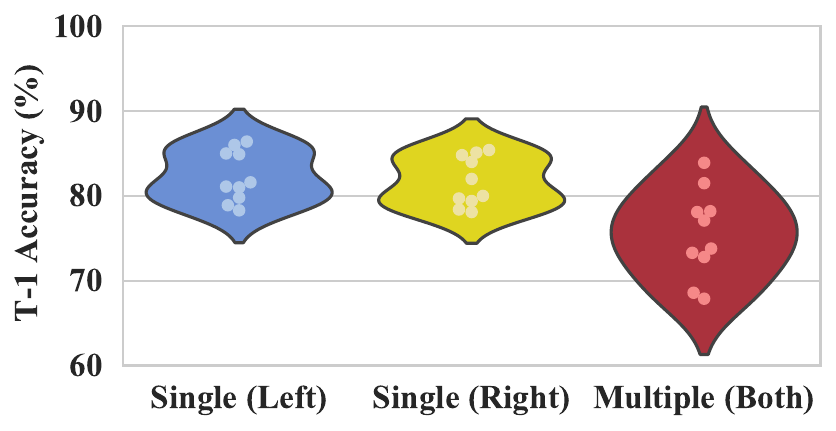}
         \caption{Word-level inference.}
         \label{fig:ir_source_word}
    \end{subfigure}
    \vspace{-0.2in}
    \caption{Performance in different IR source settings (\autoref{subsec:single_multiple_ir_source_analysis}).}
    \vspace{-0.2in}
    \label{fig:results_different_ir_sources}
\end{figure}

\subsection{Single v.s. Multiple IR Source Analysis}
\label{subsec:single_multiple_ir_source_analysis}

In \autoref{subsec:keystroke_heatmap_generation}, we discussed the method of \major{removing} image retention in considering typing both left and right controllers.
To further understand the impact of using VR controllers from a single hand or both hands, we collect data when typing at the virtual keyboard at three conditions: only left hand, only right hand, both left and right hand, and then evaluate \sysname's performance in both character-level and word-level keystroke inference.
\autoref{fig:results_different_ir_sources} shows the empirical results under the settings with a single IR source (left hand or right hand) and multiple IR sources (both hands).
Specifically, we observe that \sysname performs similarly in a single IR source but decreases approximately $5.1\%$ T-1 accuracy in character-level and $6.5\%$ T-1 accuracy in word-level keystroke inference because image \major{retention exists} when the typing interval is larger than a normal typing speed (\eg, $>$ \SI{2.0}{\second}).
Overall, \sysname maintains a \major{promising} accuracy with \major{minor variations} in recognizing virtual keystrokes from the infrared side channel under \major{multiple} input IR sources.
\looseness=-1


\section{Discussion}
\label{sec:discussion}


\subsection{Countermeasures}
\label{subsec:countermeasures}

\paragraph{Encrypted IR Transmission} With its reliance on LoS view and limited transmission of insensitive information, the current infrared (IR) transmission mechanism lacks essential security measures, such as encryption.
However, these limitations are now being challenged due to the emergence of new deployment cases in VR devices and our proposed attack.
Therefore, one potential countermeasure to defend against \sysname is redesigning the IR communication protocol to incorporate encryption, thereby
\major{modifying patterns and}
preventing eavesdropping on the transmitted \major{IR} signals in such scenarios~\cite{kim2021eavesdropping}.
Similar to other protocols like Bluetooth~\cite{padgette2017guide}, the foundation of IR encryption lies in establishing a shared key, achieved through methods such as Diffie-Hellman key exchange schemes~\cite{kallam2015diffie}, involving exchanging messages between the controllers and the VR headset to negotiate a shared secret.

\begin{figure}[t]
    \begin{subfigure}[b]{.49\linewidth}
         \centering
         \includegraphics[width=\linewidth]{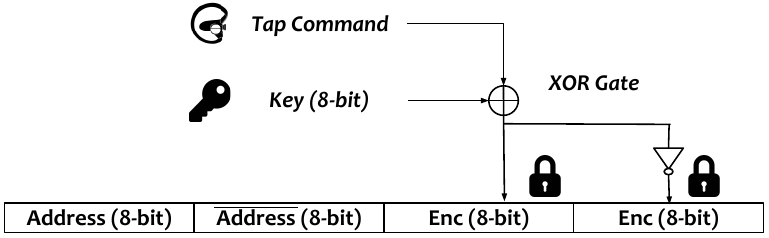}
         \vspace{-0.15in}
         \caption{\major{IR encryption scheme 1.}}
         \vspace{0.05in}
         \label{fig:ir_encryption_scheme1}
    \end{subfigure}
    \begin{subfigure}[b]{.49\linewidth}
         \centering
         \includegraphics[width=\linewidth]{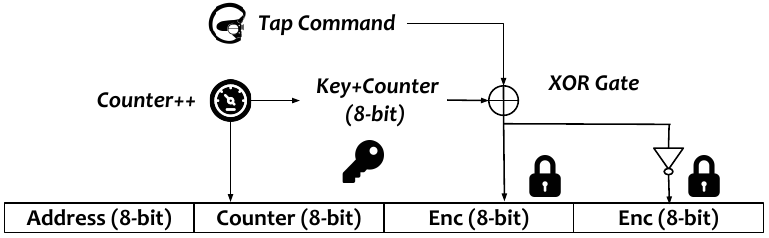}
         \vspace{-0.15in}
         \caption{\major{IR encryption scheme 2.}}
         \vspace{0.05in}
         \label{fig:ir_encryption_scheme2}
    \end{subfigure}
    \begin{subfigure}[b]{\linewidth}
         \centering
         \includegraphics[width=\linewidth]{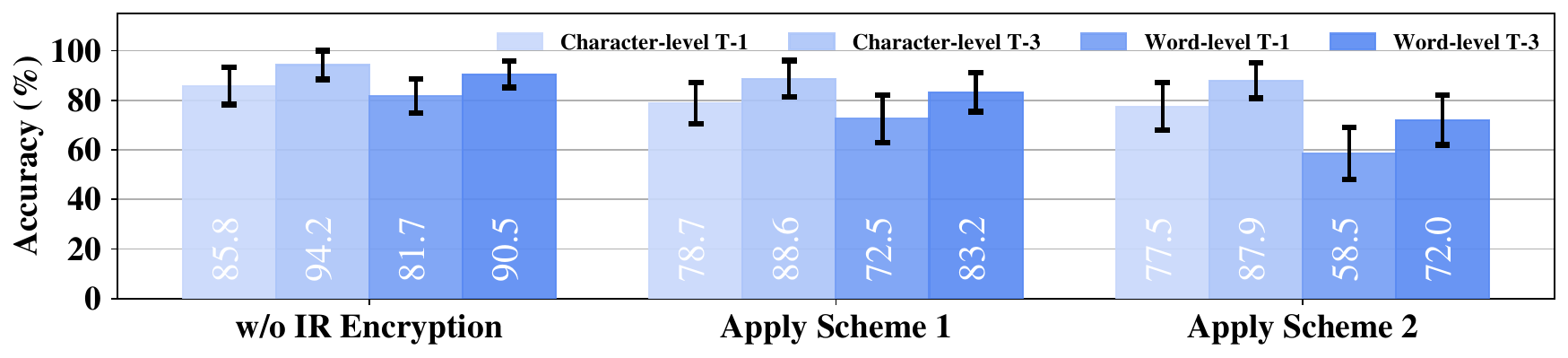}
         \vspace{-0.2in}
         \caption{\major{Empirical results in applying IR encryption schemes.}}
         \label{fig:ir_encryption_results}
    \end{subfigure}
    \vspace{-0.2in}
    \caption{\major{IR encryption schemes and defense results (\autoref{subsec:countermeasures}).}}
    \vspace{-0.2in}
    \label{fig:ir_encryption_schemes}
\end{figure}

To justify the effectiveness of IR encryption in defending against \sysname, we build upon prior work~\cite{kim2021eavesdropping} and propose two encryption schemes, illustrated in \autoref{fig:ir_encryption_scheme1} and \autoref{fig:ir_encryption_scheme2}. In both schemes, an $8$-bit key is generated by a pseudo-random number generator (pRNG)~\cite{prng}. In particular, scheme 2 includes an $8$-bit counter to encrypt the IR signals each time a VR user types on the virtual keyboard. Specifically, \autoref{eq:ir_encryption_eq} presents the two encryption schemes as:

\begin{equation}
\label{eq:ir_encryption_eq}
\small
\begin{cases}
    \text{Enc}_{S1}(\text{Tap}_{8}, \text{Key}_{8}) = \text{Tap}_{8}\oplus  \text{Key}_{8} \\ \\
    \text{Enc}_{S2}(\text{Tap}_{8}, \text{Key}_{8}) = \text{Tap}_{8}\oplus  ((\text{Key}_{8}+\text{Ctr}_{8})\;\text{mod}\;256)
\end{cases}
\vspace{0.05in}
\end{equation}

To implement these encryption schemes, we tore down Meta Oculus Quest 2 controllers and integrated them with an extra Arduino Nano MCU to encrypt the transmitted IR signals.
We then collected IR samples from the 2D sensor array in three different statutes and evaluated \sysname's performance.
\looseness=-1

\autoref{fig:ir_encryption_results} presents the empirical results. After applying the two encryption schemes, \sysname's character-level inference performance decreased by approximately $7.1\%$--$8.3\%$ (T-1) and $5.6\%$--$6.3\%$ (T-3). Notably, word-level inference performance saw a more significant drop of about $23.2\%$ (T-1) and $18.5\%$ (T-3) with scheme 2 because of the multiple encryption actions triggered by the counting process while typing long-length words.
Note that the experiments are conducted under default conditions for empirical comparison, and the countermeasures could be even more effective when combined with environmental interference in real-world scenarios.
However, the integration of extra hardware into VR controllers could impact
usability and does not guarantee security against brute-forcing attacks~\cite{yao2009initial, zhang2023evilscreen, picek2023sok} or extended attacks from \sysname if the transmitted key is extracted from the IR signals~\cite{premnath2012secret}.
\looseness=-1

\paragraph{Shuffling Virtual Keyboards} Since \sysname leverages captured IR signals from VR controllers to infer virtual keyboard input, another approach to mitigate our reported side-channel attack is to implement signal masking~\cite{batina2019csi, ni2023recovering, duan2024f2key, abuhamad2020autosen}.
That is, we can apply direct interference IR signal from other modulated light sources, \ie, fluorescent lamps~\cite{narasimhan1996effect} and TV displays~\cite{zhang2023evilscreen}, with a similar carrier frequency (\eg, \SI{38}{\kilo\hertz}~\cite{huang2023homespy}) of IR signals to obfuscate the captured signals of \sysname.
In addition, to protect \major{VR} users from keystroke inference attacks, it has been demonstrated that shuffling soft keyboards~\cite{jin2021periscope, ni2023uncovering, ni2023xporter, huang2023homespy, alghamdi2024xr, yildiran2022airtype, althebeiti2023defending} could be an effective approach, as the attacker is unable to know the randomized layout of the keyboard and cannot further infer sensitive keystrokes.
Hence, \sysname cannot effectively recover the specific keystrokes from the virtual input of a shuffled keyboard.
Nevertheless, applying signal obfuscation may interfere with the communication between the controllers and the VR headset, which impacts the link quality of the communication~\cite{ni2021simple, ni2023eavesdropping, ni2024rehsense} and further lowers functionality and usability. Likewise\major{, as demonstrated by~\cite{meteriz2022keylogging}}, shuffling virtual keyboards for each interface
\major{or after every key tap}
could increase the time spent typing on the keyboard and
\major{affect usability, especially for long keystroke input.}
\looseness=-1

\subsection{Limitations and Future Works}
\label{subsec:limitations_future_works}

Despite the \major{promising} evaluation results regarding the effectiveness of \sysname, \major{there are still} several limitations \major{in} our current research.
\major{In particular}, \sysname \major{requires to place the} IR sensor array \major{near the VR user}
to capture the \major{leaked} IR signals from the VR controllers.
However, it is worth noting that \major{\sysname can be effective in different real-world scenarios, which outperforms} other state-of-the-art non-intrusive VR \major{keystroke inference} attacks that \major{mandate placing cameras in the LoS view with sufficient light intensity and close proximity.}
Moreover, leveraging high-resolution IR cameras \major{(\eg, CoolEYE 2D module~\cite{cooleye2dmodules})} to capture thermal \major{radiations} from \major{the ambient environment} like prior \major{studies}~\cite{yu2022heatdecam} is impractical because the weak strength of IR signals emitted from VR controllers could be overwhelmed in the captured \major{thermal images}.
\looseness=-1

Additionally, our newly \major{disclosed} infrared side channel \major{provides} an orthogonal and complementary solution to other side-channel explorations in VR privacy leakages,
such as \major{unencrypted} network traffic in multi-user apps \major{or gaze information from eye-tracking sensors~\cite{wang2024gazeploit, ni2024sensor}}.
The current prototype of \sysname performs optimally when the IR sensor array is placed within a \SI{2.0}{\meter}--\SI{4.0}{\meter} range of the target VR user.
\major{This limitation} stems from the inherent constraints of the IR signals emitted by \major{commercial} VR controllers, including their limited strength and transmission \major{attenuation}.
\major{While we believe the \SI{2.0}{\meter}--\SI{4.0}{\meter} range sufficiently demonstrates the feasibility of this novel side-channel attack, enhancing the attacking distance will be a key focus of our future research efforts.}
\looseness=-1

\section{Related Works}
\label{sec:related_works}


\subsection{Keystroke Inference Attacks in VR}
\label{subsec:related_works_keystroke_inference_vr}

With the rise of Metaverse, recent studies have investigated VR attacks for stealing \major{private} information, \eg, virtual keyboard input.
Most of them exploit pre-installed malware to obtain data from built-in motion sensors (\eg, accelerometer, gyroscope) of the VR headset and train \major{multiple} models to infer keystrokes~\cite{wu2023privacy, zhang2023s, slocum2023going, ling2019know, luo2022holologger, luo2024eavesdropping}.
Moreover, Meteriz-Y{\i}ld{\i}ran~\etal~\cite{meteriz2022keylogging} presents the first keylogging attack on the virtual keyboard by placing a camera or hand tracker to monitor the VR user's hand gestures.
Similarly, Gopal~\etal proposed the \textit{Hidden Reality} attack~\cite{gopal2023hidden} that utilizes cameras to record the hand gestures of the VR user for recognizing the typing keys on the virtual keyboard.
In addition, \textit{VR-Spy}~\cite{al2021vr} also shows the Wi-Fi CSI data can be hacked to monitor hand gestures and further infer keystrokes, whereas it \major{imposes physical constraints that the user needs} to sit \major{between the transceivers}.
\looseness=-1

\major{Furthermore,} \textit{Heimdall}~\cite{luo2024eavesdropping} shows the feasibility of leveraging the \major{controllers'} button-pressing sounds to infer virtual keystrokes, which also requires the placement of a recording device at close proximity (\eg, \major{\SI{1.0}{\meter}--\SI{2.2}{\meter}}) and employs pre-trained models for inference, but its effectiveness \major{decreases} in noisy settings.
\major{Su~\etal}~\cite{su2024remote} shows that some multi-user VR apps, Rec Room, which adopts the unencrypted Photon protocol, \major{leak} the avatar's hand movement data in network traffic that \major{results in} keystroke inference attacks.
\major{One recent work, \textit{GAZEploit}~\cite{wang2024gazeploit}, exploits the avatar's gaze information recorded in online meetings to infer keystrokes typed with eye-tracking functionalities.}
Compared with these works, \sysname presents the following three advantages: \textit{(i)} it requires no malware installation and launches attacks non-intrusively at a relatively longer distance, \textit{(ii)} it \major{infers} unconstrained virtual \major{keystrokes} without training specific machine learning models for classification, and \textit{(iii)} it exploits a novel infrared side channel, which \major{discloses an orthogonal solution} with promising resilience and \major{real-world} practicality than prior attacks \major{from} other side channels.
\looseness=-1

\subsection{Other Keystroke Inference Attacks}
\label{subsec:related_works_other_keystroke_inference}

There are many efforts to exploit different side channels existing on mobile devices such as smartphones and tablets to infer \major{people's} keystrokes~\cite{wang2023beyond, spreitzer2017systematic}. 
For instance, an attacker can leverage the readings of built-in motion sensors like accelerometers, gyroscopes, and magnetometers~\cite{cai2011touchlogger, liu2015good}, system loads~\cite{yang2022eavesdropping, schwarz2018keydrown}, acoustic signals from microphones and speakers~\cite{lu2019keylistener, meteriz2022acoustictype, meteriz2021sia, chen2022swipepass}, and electromagnetic (EM) emanations from GPU~\cite{yang2022eavesdropping} or induced by human-touchscreen coupling effects~\cite{jin2021periscope} to recognize input keystrokes.
Furthermore, it is feasible to monitor changes in the channel state information (CSI) of wireless signals (\eg, Wi-Fi) to steal the typing password~\cite{yang2022wink, hu2023password}.
Besides, recent studies have demonstrated the power traces in smartphone charging processes, \ie, USB charging~\cite{ni2023xporter, su2017usb} or wireless charging~\cite{ni2023uncovering, ni2023exploiting}, exposing new attack surface for keystroke inference.
In addition, \textit{HomeSpy}~\cite{huang2023homespy} reveals that IR signals emitted from the remote control of a smart TV can be sniffed to infer input passwords and PIN codes.
Likewise, \sysname leverages the IR signals \major{leaked} from VR hand controllers to \major{infer} virtual \major{keystrokes} in an unconstrained and non-intrusive manner from the perspective of the infrared side channel that exists across devices (\eg, smart TV~\cite{huang2023homespy}) in smart home scenarios~\cite{chi2023detecting}.
\looseness=-1

\section{Conclusion}
\label{sec:conclusion}

In this paper, we present a novel side-channel attack for unconstrained keystroke inference on virtual keyboards in VR platforms by capturing the IR signals \major{leaked} from VR controllers designed for its constellation tracking system.
To validate its feasibility, we design and implement \sysname, an end-to-end attack framework that leverages a customized IR sensor array to non-intrusively capture IR signals emitted from infrared LEDs \major{embedded in} VR controllers and then recognize \major{character-level} keys and analyze typing paths to infer the \major{consecutive} keystrokes \major{within} virtual scenes.
Our extensive evaluation depicts that \sysname achieves high accuracy in recognizing keystrokes with different lengths
\major{and presents promising resilience and practicality in real-world scenarios with varying conditions.}
We hope our findings can raise public awareness of the privacy leakage from the communication characteristics between the VR headset and the VR controllers and spur research on detecting forthcoming side-channel attacks and developing new defense approaches.

\section*{Acknowledgment}

We sincerely appreciate our shepherd and all anonymous reviewers for their constructive feedback and invaluable comments.
This work was fully supported by the Research Grants Council of Hong Kong (RGC) under Grants CityU 21219223, 11218521, 11218322, R6021-20F, R1012-21, RFS2122-1S04, C2004-21G, C1029-22G, C6015-23G, N\_CityU139/21, and in part by the Innovation and Technology Commission of Hong Kong (ITC) under Mainland-Hong Kong Joint Funding Scheme (MHKJFS) MHP/135/23.
This work was also substantially supported by InnoHK initiative, The Government of the HKSAR, and Laboratory for AI-Powered Financial Technologies (AIFT).
Any opinions, findings, and conclusions in this paper are those of the authors and are not necessarily of the supported organizations.
\looseness=-1

\bibliographystyle{IEEEtran}
\bibliography{paper.bib}

\begin{thebibliography}{10}
\providecommand{\url}[1]{#1}
\csname url@samestyle\endcsname
\providecommand{\newblock}{\relax}
\providecommand{\bibinfo}[2]{#2}
\providecommand{\BIBentrySTDinterwordspacing}{\spaceskip=0pt\relax}
\providecommand{\BIBentryALTinterwordstretchfactor}{4}
\providecommand{\BIBentryALTinterwordspacing}{\spaceskip=\fontdimen2\font plus
\BIBentryALTinterwordstretchfactor\fontdimen3\font minus \fontdimen4\font\relax}
\providecommand{\BIBforeignlanguage}[2]{{%
\expandafter\ifx\csname l@#1\endcsname\relax
\typeout{** WARNING: IEEEtran.bst: No hyphenation pattern has been}%
\typeout{** loaded for the language `#1'. Using the pattern for}%
\typeout{** the default language instead.}%
\else
\language=\csname l@#1\endcsname
\fi
#2}}
\providecommand{\BIBdecl}{\relax}
\BIBdecl

\bibitem{slocum2023going}
C.~Slocum, Y.~Zhang, N.~Abu-Ghazaleh, and J.~Chen, ``Going through the motions: {AR/VR} keylogging from user head motions,'' in \emph{Proceedings of the 32nd USENIX Security Symposium}, 2023.

\bibitem{zhang2023s}
Y.~Zhang, C.~Slocum, J.~Chen, and N.~Abu-Ghazaleh, ``It’s all in your head (set): Side-channel attacks on {AR/VR} systems,'' in \emph{Proceedings of the 32nd USENIX Security Symposium}, 2023.

\bibitem{wu2023privacy}
Y.~Wu, C.~Shi, T.~Zhang, P.~Walker, J.~Liu, N.~Saxena, and Y.~Chen, ``Privacy leakage via unrestricted motion-position sensors in the age of virtual reality: A study of snooping typed input on virtual keyboards,'' in \emph{Proceedings of the IEEE Symposium on Security and Privacy (SP)}, 2023.

\bibitem{luo2022holologger}
S.~Luo, X.~Hu, and Z.~Yan, ``Holologger: Keystroke inference on mixed reality head-mounted displays,'' in \emph{Proceedings of the IEEE Conference on Virtual Reality and 3D User Interfaces (VR)}, 2022.

\bibitem{al2021vr}
A.~Al~Arafat, Z.~Guo, and A.~Awad, ``{VR-Spy}: A side-channel attack on virtual key-logging in {VR} headsets,'' in \emph{Proceedings of the IEEE Conference on Virtual Reality and 3D User Interfaces (VR)}, 2021.

\bibitem{meteriz2022keylogging}
{\"U}.~Meteriz-Y{\i}ld{\i}ran, N.~F. Y{\i}ld{\i}ran, A.~Awad, and D.~Mohaisen, ``A keylogging inference attack on air-tapping keyboards in virtual environments,'' in \emph{Proceedings of the IEEE Conference on Virtual Reality and 3D User Interfaces (VR)}, 2022.

\bibitem{su2024remote}
Z.~Su, K.~Cai, R.~Beeler, L.~Dresel, A.~Garcia, I.~Grishchenko, Y.~Tian, C.~Kruegel, and G.~Vigna, ``Remote keylogging attacks in multi-user vr applications,'' \emph{arXiv preprint arXiv:2405.14036}, 2024.

\bibitem{wang2024gazeploit}
H.~Wang, Z.~Zhan, H.~Shan, S.~Dai, M.~Panoff, and S.~Wang, ``{GAZEploit}: Remote keystroke inference attack by gaze estimation from avatar views in {VR/MR} devices,'' in \emph{Proceedings of the ACM SIGSAC Conference on Computer and Communications Security (CCS)}, 2024.

\bibitem{gopal2023hidden}
S.~R.~K. Gopal, D.~Shukla, J.~D. Wheelock, and N.~Saxena, ``{Hidden Reality}: Caution, your hand gesture inputs in the immersive virtual world are visible to all!'' in \emph{Proceedings of the 32nd USENIX Security Symposium}, 2023.

\bibitem{luo2024eavesdropping}
S.~Luo, A.~Nguyen, H.~Farooq, K.~Sun, and Z.~Yan, ``Eavesdropping on controller acoustic emanation for keystroke inference attack in virtual reality,'' in \emph{Proceedings of the Network and Distributed System Security (NDSS) Symposium}, 2024.

\bibitem{ling2019know}
Z.~Ling, Z.~Li, C.~Chen, J.~Luo, W.~Yu, and X.~Fu, ``I know what you enter on gear {VR},'' in \emph{Proceedings of the IEEE Conference on Communications and Network Security (CNS)}, 2019.

\bibitem{khalili2024virtual}
H.~Khalili, A.~Chen, T.~Papaiakovou, T.~Jacques, H.-J. Chien, C.~Liu, A.~Ding, A.~Hass, S.~Zonouz, and N.~Sehatbakhsh, ``Virtual keymysteries unveiled: Detecting keystrokes in {VR} with external side-channels,'' in \emph{Proceedings of the IEEE Security and Privacy Workshops (SPW)}, 2024.

\bibitem{outsideinvsinsideout}
Pimax, ``Pose tracking methods: Outside-in vs inside-out tracking in {VR},'' \url{https://pimax.com/pose-tracking-methods-outside-in-vs-inside-out-tracking-in-vr}, 2023.

\bibitem{vrcontroller}
D.~Gajsek, ``Vr controllers: The way of interacting with the virtual worlds,'' \url{https://circuitstream.com/blog/vr-controllers-the-way-of-interacting-with-the-virtual-worlds}, 2022.

\bibitem{htcvivepro}
{HTC VIVE}, ``Vive pro,'' \url{https://www.vive.com/eu/product/vive-pro}, 2023.

\bibitem{oculusquest2}
Meta, ``{Meta Quest 2}: Immersive all-in-one {VR} headset,'' \url{https://www.meta.com/quest/products/quest-2}, 2023.

\bibitem{gerloni2018immersive}
I.~G. Gerloni, V.~Carchiolo, F.~R. Vitello, E.~Sciacca, U.~Becciani, A.~Costa, S.~Riggi, F.~L. Bonali, E.~Russo, L.~Fallati \emph{et~al.}, ``Immersive virtual reality for earth sciences,'' in \emph{Proceedings of the Federated Conference on Computer Science and Information Systems (FedCSIS)}, 2018.

\bibitem{pico4}
PICO, ``Live the game with {PICO} 4 all-in-one {VR} headset,'' \url{https://www.picoxr.com/global/products/pico4}, 2023.

\bibitem{hpreverb}
W.~Greenwald, ``{HP Reverb G2} review,'' \url{https://www.pcmag.com/reviews/hp-reverb-g2}, 2023.

\bibitem{ps5nextgenvr}
{PlayStation Blog}, ``Next-gen {VR} on {PS5}: the new controller,'' \url{https://blog.playstation.com/2021/03/18/next-gen-vr-on-ps5-the-new-controller/}, 2021.

\bibitem{tsop1838}
{Twins Chip}, ``{TSOP1838} infrared sensor,'' \url{https://www.twinschip.com/TSOP1838_Infrared_Sensor}, 2023.

\bibitem{daud2013application}
S.~Daud, S.~M. Sobani, M.~Ramiee, N.~Mahmood, P.~Leow, and F.~C. Harun, ``Application of infrared sensor for shape detection,'' in \emph{Proceedings of the IEEE International Conference on Photonics (ICP)}, 2013.

\bibitem{onewayfilm}
{Coavas Store}, ``{Coavas One Way Privacy Window Film},'' 2024, \url{https://www.amazon.com/Coavas-Privacy-Window-Film-Tools/dp/B0CLLWX5MS}.

\bibitem{huang2023homespy}
K.~Huang, Y.~Zhou, K.~Zhang, J.~Xu, J.~Chen, D.~Tang, and K.~Zhang, ``{HOMESPY}: The invisible sniffer of infrared remote control of smart {TVs},'' in \emph{Proceedings of the 32nd USENIX Security Symposium}, 2023.

\bibitem{visionpro}
{Apple Inc.}, ``{Vision Pro},'' \url{https://www.apple.com/apple-vision-pro}, 2023.

\bibitem{irsensorswebsite}
Elecbee, ``{1838T} infrared sensor receiver module board remote controller {IR} sensor with cable for arduino,'' \url{https://www.elecbee.com/en-26475-10pcs-1838T-Infrared-Sensor-Receiver-Module-Board-Remote-Controller-IR-Sensor-with-Cable-for-Arduino-products-that-work-with-official-Arduino-boards}, 2023.

\bibitem{tsop1838tsensor}
{Open Impulse}, ``Tl1838 infrared receiver datasheet,'' \url{http://eeshop.unl.edu/pdf/VS1838-Infrared-Receiver-datasheet.pdf}, 2023.

\bibitem{benet2002using}
G.~Benet, F.~Blanes, J.~E. Sim{\'o}, and P.~P{\'e}rez, ``Using infrared sensors for distance measurement in mobile robots,'' \emph{Robotics and autonomous systems}, 2002.

\bibitem{uyanik2013study}
G.~K. Uyan{\i}k and N.~G{\"u}ler, ``A study on multiple linear regression analysis,'' \emph{Procedia-Social and Behavioral Sciences}, vol. 106, pp. 234--240, 2013.

\bibitem{yang2022eavesdropping}
B.~Yang, R.~Chen, K.~Huang, J.~Yang, and W.~Gao, ``Eavesdropping user credentials via gpu side channels on smartphones,'' in \emph{Proceedings of the 27th ACM International Conference on Architectural Support for Programming Languages and Operating Systems (ASPLOS)}, 2022.

\bibitem{zhao2023odam}
C.~Zhao and A.~B. Chan, ``Odam: Gradient-based instance-specific visual explanations for object detection,'' in \emph{Proceedings of the 11th International Conference on Learning Representations (ICLR)}, 2023.

\bibitem{jin2021periscope}
W.~Jin, S.~Murali, H.~Zhu, and M.~Li, ``Periscope: A keystroke inference attack using human coupled electromagnetic emanations,'' in \emph{Proceedings of the ACM SIGSAC Conference on Computer and Communications Security (CCS)}, 2021.

\bibitem{cronin2021charger}
P.~Cronin, X.~Gao, C.~Yang, and H.~Wang, ``{Charger-Surfing}: Exploiting a power line side-channel for smartphone information leakage,'' in \emph{Proceedings of the 30th USENIX Security Symposium}, 2021.

\bibitem{boundingbox}
MathWorks, ``Bounding box of polyshape,'' \url{https://www.mathworks.com/help/matlab/ref/polyshape.boundingbox.html}, 2023.

\bibitem{wu2023chatgpt}
H.~Wu, W.~Wang, Y.~Wan, W.~Jiao, and M.~Lyu, ``{ChatGPT} or grammarly? evaluating {ChatGPT} on grammatical error correction benchmark,'' \emph{arXiv preprint arXiv:2303.13648}, 2023.

\bibitem{penteado2023evaluating}
M.~C. Penteado and F.~Perez, ``Evaluating {GPT-3.5} and {GPT-4} on grammatical error correction for brazilian portuguese,'' \emph{arXiv preprint arXiv:2306.15788}, 2023.

\bibitem{saeidnia2023welcome}
H.~R. Saeidnia, ``Welcome to the gemini era: Google deepmind and the information industry,'' \emph{Library Hi Tech News}, no. ahead-of-print, 2023.

\bibitem{cambridgeenglish}
{The University of Cambridge}, ``{Cambridge English Vocabulary List},'' 2012, \url{https://www.cambridgeenglish.org/images/84669-pet-vocabulary-list.pdf}.

\bibitem{sfh4055}
{OSRAM Opto Semiconductors}, ``Sfh 4055,'' \url{https://look.ams-osram.com/m/7723eb0dcefc7b33/original/SFH-4055.pdf}, 2019.

\bibitem{tsal6400}
{Vishay Semiconductors}, ``High power infrared emitting diode,'' \url{https://www.vishay.com/docs/81011/tsal6400.pdf}, 2014.

\bibitem{ir333a}
{Everlight Europe}, ``{IR333-A},'' \url{https://everlighteurope.com/ir-emitters/159/IR333A.html}, 2023.

\bibitem{hu2023password}
J.~Hu, H.~Wang, T.~Zheng, J.~Hu, Z.~Chen, H.~Jiang, and J.~Luo, ``Password-stealing without hacking: {Wi-Fi} enabled practical keystroke eavesdropping,'' in \emph{Proceedings of the ACM SIGSAC Conference on Computer and Communications Security (CCS)}, 2023.

\bibitem{kim2021eavesdropping}
M.~Kim and T.~Suh, ``Eavesdropping vulnerability and countermeasure in infrared communication for iot devices,'' \emph{Sensors}, 2021.

\bibitem{padgette2017guide}
J.~Padgette, K.~Scarfone, and L.~Chen, ``Guide to bluetooth security,'' \emph{NIST special publication}, 2017.

\bibitem{kallam2015diffie}
S.~Kallam, ``Diffie-hellman: key exchange and public key cryptosystems,'' \emph{Master degree of Science, Math and Computer Science, Department of India State University, USA}, 2015.

\bibitem{prng}
{Leonardo Miliani}, ``{pRNG.h: pretty Random Number Generator},'' 2016, \url{https://github.com/leomil72/pRNG}.

\bibitem{yao2009initial}
T.~Yao, K.~Fukui, J.~Nakashima, and T.~Nakai, ``Initial common secret key sharing using random plaintexts for short-range wireless communications,'' \emph{IEEE Transactions on Consumer Electronics}, 2009.

\bibitem{zhang2023evilscreen}
Y.~Zhang, S.~Ma, T.~Chen, J.~Li, R.~H. Deng, and E.~Bertino, ``Evilscreen attack: Smart {TV} hijacking via multi-channel remote control mimicry,'' \emph{IEEE Transactions on Dependable and Secure Computing (TDSC)}, 2023.

\bibitem{picek2023sok}
S.~Picek, G.~Perin, L.~Mariot, L.~Wu, and L.~Batina, ``Sok: Deep learning-based physical side-channel analysis,'' \emph{ACM Computing Surveys}, 2023.

\bibitem{premnath2012secret}
S.~N. Premnath, S.~Jana, J.~Croft, P.~L. Gowda, M.~Clark, S.~K. Kasera, N.~Patwari, and S.~V. Krishnamurthy, ``Secret key extraction from wireless signal strength in real environments,'' \emph{IEEE Transactions on Mobile Computing (TMC)}, 2012.

\bibitem{batina2019csi}
L.~Batina, S.~Bhasin, D.~Jap, and S.~Picek, ``{CSI NN}: Reverse engineering of neural network architectures through electromagnetic side channel,'' in \emph{Proceedings of the USENIX Security Symposium}, 2019.

\bibitem{ni2023recovering}
T.~Ni, X.~Zhang, and Q.~Zhao, ``Recovering fingerprints from in-display fingerprint sensors via electromagnetic side channel,'' in \emph{Proceedings of the ACM SIGSAC Conference on Computer and Communications Security (CCS)}, 2023.

\bibitem{duan2024f2key}
D.~Duan, Z.~Sun, T.~Ni, S.~Li, X.~Jia, W.~Xu, and T.~Li, ``F2key: Dynamically converting your face into a private key based on cots headphones for reliable voice interaction,'' in \emph{Proceedings of the 22nd Annual International Conference on Mobile Systems, Applications and Services (MobiSys)}, 2024.

\bibitem{abuhamad2020autosen}
M.~Abuhamad, T.~Abuhmed, D.~Mohaisen, and D.~Nyang, ``Autosen: Deep-learning-based implicit continuous authentication using smartphone sensors,'' \emph{IEEE Internet of Things Journal (IoTJ)}, 2020.

\bibitem{narasimhan1996effect}
R.~Narasimhan, M.~D. Audeh, and J.~M. Kahn, ``Effect of electronic-ballast fluorescent lighting on wireless infrared links,'' \emph{IEE Proceedings-Optoelectronics}, 1996.

\bibitem{ni2023uncovering}
T.~Ni, X.~Zhang, C.~Zuo, J.~Li, Z.~Yan, W.~Wang, W.~Xu, X.~Luo, and Q.~Zhao, ``Uncovering user interactions on smartphones via contactless wireless charging side channels,'' in \emph{Proceedings of the IEEE Symposium on Security and Privacy (SP)}, 2023.

\bibitem{ni2023xporter}
T.~Ni, Y.~Chen, W.~Xu, L.~Xue, and Q.~Zhao, ``Xporter: A study of the multi-port charger security on privacy leakage and voice injection,'' in \emph{Proceedings of the 29th Annual International Conference on Mobile Computing and Networking (MobiCom)}, 2023.

\bibitem{alghamdi2024xr}
A.~Alghamdi, A.~Alkinoon, A.~Alghuried, and D.~Mohaisen, ``xr-droid: A benchmark dataset for ar/vr and security applications,'' \emph{IEEE Transactions on Dependable and Secure Computing (TDSC)}, 2024.

\bibitem{yildiran2022airtype}
N.~F. Y{\i}ld{\i}ran, {\"U}.~Meteriz-Yildiran, and D.~Mohaisen, ``Airtype: an air-tapping keyboard for augmented reality environments,'' in \emph{Proceedings of the IEEE Conference on Virtual Reality and 3D User Interfaces Abstracts and Workshops (VRW)}, 2022.

\bibitem{althebeiti2023defending}
H.~Althebeiti, R.~Gedawy, A.~Alghuried, D.~Nyang, and D.~Mohaisen, ``Defending airtype against inference attacks using 3d in-air keyboard layouts: Design and evaluation,'' in \emph{Proceedings of the International Conference on Information Security Applications (ICISA)}, 2023.

\bibitem{ni2021simple}
T.~Ni, Y.~Chen, K.~Song, and W.~Xu, ``A simple and fast human activity recognition system using radio frequency energy harvesting,'' in \emph{Adjunct Proceedings of the 2021 ACM International Joint Conference on Pervasive and Ubiquitous Computing and Proceedings of the 2021 ACM International Symposium on Wearable Computers (UbiComp-CPD)}, 2021.

\bibitem{ni2023eavesdropping}
T.~Ni, G.~Lan, J.~Wang, Q.~Zhao, and W.~Xu, ``Eavesdropping mobile app activity via radio-frequency energy harvesting,'' in \emph{Proceedings of the 32nd USENIX Security Symposium}, 2023.

\bibitem{ni2024rehsense}
T.~Ni, Z.~Sun, M.~Han, Y.~Xie, G.~Lan, Z.~Li, T.~Gu, and W.~Xu, ``Rehsense: Towards battery-free wireless sensing via radio frequency energy harvesting,'' in \emph{Proceedings of the 25th International Symposium on Theory, Algorithmic Foundations, and Protocol Design for Mobile Networks and Mobile Computing (MobiHoc)}, 2024.

\bibitem{cooleye2dmodules}
E.~Technologies, ``{CoolEYE IR 2D Modules},'' 2020, \url{https://www.excelitas.com/product-category/cooleye-ir-2d-modules}.

\bibitem{yu2022heatdecam}
Z.~Yu, Z.~Li, Y.~Chang, S.~Fong, J.~Liu, and N.~Zhang, ``Heatdecam: detecting hidden spy cameras via thermal emissions,'' in \emph{Proceedings of the ACM SIGSAC Conference on Computer and Communications Security (CCS)}, 2022.

\bibitem{ni2024sensor}
T.~Ni, ``Sensor security in virtual reality: Exploration and mitigation,'' in \emph{Proceedings of the 22nd Annual International Conference on Mobile Systems, Applications and Services (MobiSys)}, 2024.

\bibitem{wang2023beyond}
G.~Wang, C.~Zhou, Y.~Wang, B.~Chen, H.~Guo, and Q.~Yan, ``Beyond boundaries: A comprehensive survey of transferable attacks on ai systems,'' \emph{arXiv preprint arXiv:2311.11796}, 2023.

\bibitem{spreitzer2017systematic}
R.~Spreitzer, V.~Moonsamy, T.~Korak, and S.~Mangard, ``Systematic classification of side-channel attacks: A case study for mobile devices,'' \emph{IEEE Communications Surveys and Tutorials}, 2017.

\bibitem{cai2011touchlogger}
L.~Cai and H.~Chen, ``Touchlogger: Inferring keystrokes on touch screen from smartphone motion,'' in \emph{Proceedings of the 6th USENIX Workshop on Hot Topics in Security (HotSec)}, 2011.

\bibitem{liu2015good}
X.~Liu, Z.~Zhou, W.~Diao, Z.~Li, and K.~Zhang, ``When good becomes evil: Keystroke inference with smartwatch,'' in \emph{Proceedings of the ACM SIGSAC Conference on Computer and Communications Security (CCS)}, 2015.

\bibitem{schwarz2018keydrown}
M.~Schwarz, M.~Lipp, D.~Gruss, S.~Weiser, C.~L.~N. Maurice, R.~Spreitzer, and S.~Mangard, ``Keydrown: Eliminating software-based keystroke timing side-channel attacks,'' in \emph{Proceedings of the Network and Distributed System Security Symposium (NDSS)}, 2018.

\bibitem{lu2019keylistener}
L.~Lu, J.~Yu, Y.~Chen, Y.~Zhu, X.~Xu, G.~Xue, and M.~Li, ``Keylistener: Inferring keystrokes on {QWERTY} keyboard of touch screen through acoustic signals,'' in \emph{Proceedings of the IEEE Conference on Computer Communications (INFOCOM)}, 2019.

\bibitem{meteriz2022acoustictype}
{\"U}.~Meteriz~Y{\`y}ld{\`y}ran, N.~F. Y{\`y}ld{\`y}ran, and D.~Mohaisen, ``Acoustictype: Smartwatch-enabled cross-device text entry method using keyboard acoustics,'' in \emph{Proceedings of the CHI Conference on Human Factors in Computing Systems Extended Abstracts}, 2022.

\bibitem{meteriz2021sia}
{\"U}.~Meteriz-Y{\i}ld{\i}ran, N.~F. Yildiran, and D.~Mohaisen, ``Sia: Smartwatch-enabled inference attacks on physical keyboards using acoustic signals,'' in \emph{Proceedings of the 20th Workshop on Workshop on Privacy in the Electronic Society}, 2021.

\bibitem{chen2022swipepass}
Y.~Chen, T.~Ni, W.~Xu, and T.~Gu, ``Swipepass: Acoustic-based second-factor user authentication for smartphones,'' \emph{Proceedings of the ACM on Interactive, Mobile, Wearable and Ubiquitous Technologies (IMWUT)}, 2022.

\bibitem{yang2022wink}
E.~Yang, Q.~He, and S.~Fang, ``{WINK}: Wireless inference of numerical keystrokes via zero-training spatiotemporal analysis,'' in \emph{Proceedings of the ACM SIGSAC Conference on Computer and Communications Security (CCS)}, 2022.

\bibitem{su2017usb}
Y.~Su, D.~Genkin, D.~Ranasinghe, and Y.~Yarom, ``{USB} snooping made easy: crosstalk leakage attacks on {USB} hubs,'' in \emph{Proceedings of the 26th USENIX Security Symposium}, 2017.

\bibitem{ni2023exploiting}
T.~Ni, J.~Li, X.~Zhang, C.~Zuo, W.~Wang, W.~Xu, X.~Luo, and Q.~Zhao, ``Exploiting contactless side channels in wireless charging power banks for user privacy inference via few-shot learning,'' in \emph{Proceedings of the 29th Annual International Conference on Mobile Computing and Networking (MobiCom)}, 2023.

\bibitem{chi2023detecting}
H.~Chi, Q.~Zeng, and X.~Du, ``Detecting and handling {IoT} interaction threats in multi-platform multi-control-channel smart homes,'' in \emph{Proceedings of the USENIX Security Symposium}, 2023.

\bibitem{neamoniti2022hand}
S.~Neamoniti and V.~Kasapakis, ``Hand tracking vs motion controllers: The effects on immersive virtual reality game experience,'' in \emph{Proceedings of the IEEE International Symposium on Multimedia (ISM)}, 2022.

\bibitem{zhu2020blinkey}
H.~Zhu, W.~Jin, M.~Xiao, S.~Murali, and M.~Li, ``Blinkey: A two-factor user authentication method for virtual reality devices,'' \emph{Proceedings of the ACM on Interactive, Mobile, Wearable and Ubiquitous Technologies (IMWUT)}, 2020.

\bibitem{zhu2023soundlock}
H.~Zhu, M.~Xiao, D.~Sherman, and M.~Li, ``Soundlock: A novel user authentication scheme for vr devices using auditory-pupillary response.'' in \emph{Proceedings of the Network and Distributed System Security Symposium (NDSS)}, 2023.

\end{thebibliography}

\appendix

\subsection{Outlook of IR Sensor Array Prototype}

\begin{figure}[ht]
    \centering
    \includegraphics[width=.96\linewidth]{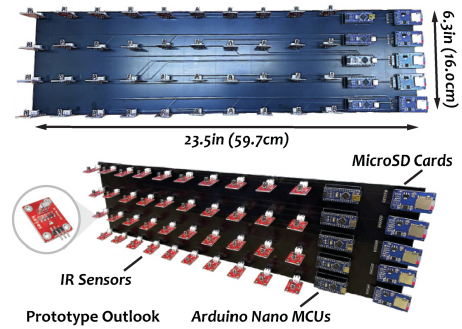}
    \caption{Prototype of the customized 2D IR sensor array (\autoref{subsec:customized_ir_sensor_array}), which consists of $40$ 1838T IR sensors~\cite{irsensorswebsite}, five Arduino Nano MCUs, and five MicroSD card adapters.}
    \vspace{-0.1in}
    \label{fig:customized_ir_sensor_array}
\end{figure}



\begin{figure}[t]
    \begin{subfigure}[b]{.495\linewidth}
         \centering
         \includegraphics[width=\linewidth]{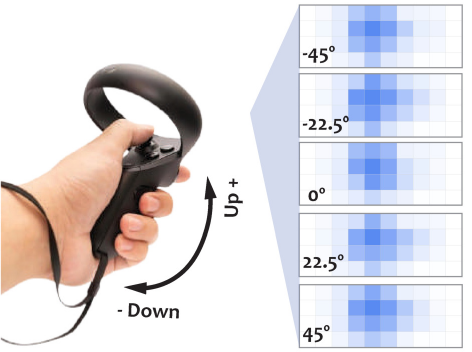}
         \caption{Up-to-down arm rotation.}
         \label{fig:updown_arm_rotation}
    \end{subfigure}
    \begin{subfigure}[b]{.495\linewidth}
         \centering
         \includegraphics[width=\linewidth]{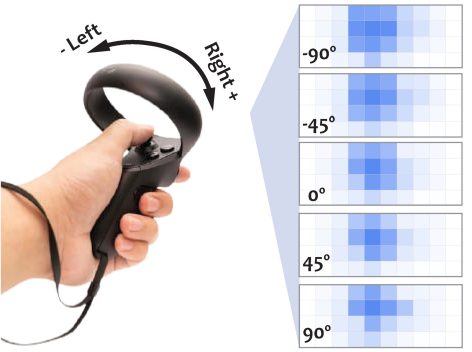}
         \caption{Left-to-right arm rotation.}
         \label{fig:leftright_arm_rotation}
    \end{subfigure}
    \vspace{-0.15in}
    \caption{Controller omnidirectional analysis (\autoref{subsec:controller_omnidirections_analysis}).}
    \vspace{-0.2in}
    \label{fig:controller_omnidirectional_analysis}
\end{figure}

\subsection{Controller Omnidirectional Analysis}
\label{subsec:controller_omnidirections_analysis}

As discussed in \autoref{subsec:infrared_led_vr}, multiple infrared LEDs are embedded around the ring of the VR controller, resulting in an omnidirectional IR radiation pattern. This means that IR signals emitted from the VR controllers disperse in all directions when the user unconsciously rotates their arm, potentially causing additional interference with the IR sensor array. To investigate the effects of omnidirectional radiation, we instructed \major{participants} to type the same key on a virtual keyboard at various wrist orientation angles. \autoref{fig:updown_arm_rotation} and \autoref{fig:leftright_arm_rotation} illustrate the heatmaps generated when the user holds the VR controller to type the key “G” at up-to-down orientation angles ranging from $-45^{\circ}$ to $45^{\circ}$ and left-to-right orientation angles ranging from $-90^{\circ}$ to $90^{\circ}$, respectively.
We observe that, although there are slight variations in the generated IR heatmaps, the densest color regions consistently focus on the same key on the virtual keyboard. This indicates that the omnidirectional emissions of IR signals from the VR controller have a limited impact on \sysname's keystroke inference performance. Given the limited power of the embedded LEDs and the substantial distance between the VR controllers and the IR sensor array, multiple LEDs on a VR controller can be effectively summarized as a single IR source. This is because the majority of the IR signals' power from these LEDs is directed towards the front, pointing towards the target virtual key, thereby leading to keystroke leakage from the infrared side channel. Note that we considered the up-to-down and left-to-right orientation angles for the \major{participants'} hands under typical keyboard typing conditions.

\subsection{Controller-based v.s. Controller-less VR}
\label{subsec:discussion_apple_vision_pro}

Recently, Apple released its highly-anticipated VR headset, the Apple Vision Pro~\cite{visionpro} in February 2024.
Unlike traditional VR devices, the Apple Vision Pro selects a controller-less design, revolutionizing user interaction by incorporating hand-tracking technology.
On the other hand, other VR devices like the Meta Oculus Quest 2 and PICO 4 All-in-One have already supported native hand tracking~\cite{neamoniti2022hand}. Consequently, when \major{VR} users rely on hand-tracking mode, there is no emission of IR signals, rendering \sysname ineffective in such scenarios.
Nevertheless, we have investigated $10$ popular VR devices from different metrics, and \autoref{tab:investigation_vr_devices} show that most VR devices still adopt hand controllers to enhance user interactions within virtual environments for several reasons:

\begin{packeditemize}

\item \textbf{Lack Haptic Feedback:} Compared to tactile feedback from pressing physical buttons on VR controllers, using hand-tracking mode to interact with the VR headset often lacks haptic feedback, resulting in a higher false positive rate.

\item \textbf{High Response Latency in Hand-tracking Mode:} The response latency experienced in the VR user's interactions with hand-tracking cannot match the high responsiveness of other controller-based VR interactions~\cite{vrcontroller}.

\item \textbf{Limited Availability of Hand-tracking Scenarios:} There are a limited number of commodity VR apps that currently support the hand-tracking mode, which restricts VR apps' generalization across different VR platforms.

\item \textbf{Threats of Potential Replay Attacks:} Using embedded cameras to track hand gestures is susceptible to imitation replay attacks~\cite{zhu2020blinkey, zhu2023soundlock} because human gestures are easy to be monitored, simulated and replicated by adversaries.

\end{packeditemize}

As such, we can observe that the majority of mainstream VR devices, including Meta Oculus Quest, PICO, and HTC VIVE, continue to incorporate hand controllers in their products and over $50\%$ of them adopt the IR-based constellation tracking systems, making \sysname an applicable attack.
\looseness=-1

\begin{table}[ht]
\centering
\scriptsize
\setlength{\tabcolsep}{3pt}
\renewcommand{\arraystretch}{1.2}
\begin{tabular}{cccc}
\hline
\headcol\multicolumn{1}{!{\vrule width 0.5pt}c!{\vrule width 0.5pt}}{\textcolor{white}{\textbf{VR Devices}}}  & \multicolumn{1}{c!{\vrule width 0.5pt}}{\textcolor{white}{\textbf{Hand Controllers?}}} & \multicolumn{1}{c!{\vrule width 0.5pt}}{\textcolor{white}{\textbf{Constellation?}}} & \multicolumn{1}{c!{\vrule width 0.5pt}}{\textcolor{white}{\textbf{Hand-tracking?}}} \\ \hline
\multicolumn{1}{!{\vrule width 0.5pt}c!{\vrule width 0.5pt}}{Meta Oculus Quest 2} & \multicolumn{1}{c!{\vrule width 0.5pt}}{\myYescirc} & \multicolumn{1}{c!{\vrule width 0.5pt}}{\myYescirc} & \multicolumn{1}{c!{\vrule width 0.5pt}}{\myYescirc}                             \\ \hline
\multicolumn{1}{!{\vrule width 0.5pt}c!{\vrule width 0.5pt}}{PICO 4 All-in-One} & \multicolumn{1}{c!{\vrule width 0.5pt}}{\myYescirc} & \multicolumn{1}{c!{\vrule width 0.5pt}}{\myYescirc}  & \multicolumn{1}{c!{\vrule width 0.5pt}}{\myYescirc}                             \\ \hline
\multicolumn{1}{!{\vrule width 0.5pt}c!{\vrule width 0.5pt}}{HTC Vive Pro 2} & \multicolumn{1}{c!{\vrule width 0.5pt}}{\myYescirc} & \multicolumn{1}{c!{\vrule width 0.5pt}}{\myYescirc}  & \multicolumn{1}{c!{\vrule width 0.5pt}}{\myYescirc}                             \\ \hline
\multicolumn{1}{!{\vrule width 0.5pt}c!{\vrule width 0.5pt}}{Sony PlayStation VR 2} & \multicolumn{1}{c!{\vrule width 0.5pt}}{\myYescirc}  & \multicolumn{1}{c!{\vrule width 0.5pt}}{\myNocirc} & \multicolumn{1}{c!{\vrule width 0.5pt}}{\myNocirc}                             \\ \hline
\multicolumn{1}{!{\vrule width 0.5pt}c!{\vrule width 0.5pt}}{Meta Oculus Quest Pro} & \multicolumn{1}{c!{\vrule width 0.5pt}}{\myYescirc} & \multicolumn{1}{c!{\vrule width 0.5pt}}{\myYescirc}  & \multicolumn{1}{c!{\vrule width 0.5pt}}{\myYescirc}                             \\ \hline
\multicolumn{1}{!{\vrule width 0.5pt}c!{\vrule width 0.5pt}}{Meta Oculus Quest 3} & \multicolumn{1}{c!{\vrule width 0.5pt}}{\myYescirc} & \multicolumn{1}{c!{\vrule width 0.5pt}}{\myYescirc}  & \multicolumn{1}{c!{\vrule width 0.5pt}}{\myYescirc}                             \\ \hline
\multicolumn{1}{!{\vrule width 0.5pt}c!{\vrule width 0.5pt}}{Valve Index} & \multicolumn{1}{c!{\vrule width 0.5pt}}{\myYescirc} & \multicolumn{1}{c!{\vrule width 0.5pt}}{\myYescirc}  & \multicolumn{1}{c!{\vrule width 0.5pt}}{\myNocirc}                             \\ \hline
\multicolumn{1}{!{\vrule width 0.5pt}c!{\vrule width 0.5pt}}{HP WMR Headset} & \multicolumn{1}{c!{\vrule width 0.5pt}}{\myYescirc} & \multicolumn{1}{c!{\vrule width 0.5pt}}{\myYescirc}  & \multicolumn{1}{c!{\vrule width 0.5pt}}{\myNocirc}                             \\ \hline
\multicolumn{1}{!{\vrule width 0.5pt}c!{\vrule width 0.5pt}}{Dell Visor} & \multicolumn{1}{c!{\vrule width 0.5pt}}{\myYescirc} & \multicolumn{1}{c!{\vrule width 0.5pt}}{\myNocirc}  & \multicolumn{1}{c!{\vrule width 0.5pt}}{\myNocirc}                             \\ \hline
\multicolumn{1}{!{\vrule width 0.5pt}c!{\vrule width 0.5pt}}{Apple Vision Pro} & \multicolumn{1}{c!{\vrule width 0.5pt}}{\myNocirc} & \multicolumn{1}{c!{\vrule width 0.5pt}}{\myNocirc}  & \multicolumn{1}{c!{\vrule width 0.5pt}}{\myYescirc}                             \\ \hline
\end{tabular}%
\caption{\major{Investigation results of 10 popular VR devices.}}
\label{tab:investigation_vr_devices}
\vspace{-0.25in}
\end{table}

\end{document}